\documentclass[
twocolumn,
prl,
]{revtex4-1}
\usepackage{graphicx}
\usepackage{dcolumn}
\usepackage{bm}
\usepackage{epstopdf}

\begin{document}

\preprint{APS/123-QED}

\title{Quasi-particle evidence for the nematic state above $T_{\rm{c}}$ in Sr$_x$Bi$_2$Se$_3$}

\author{Yue Sun,$^{1,6}$}
\email{sunyue@phys.aoyama.ac.jp}
\author{Shunichiro Kittaka,$^1$ Toshiro Sakakibara,$^1$ Kazushige Machida,$^2$ Jinghui Wang,$^3$ Jinsheng Wen$^3$, Xiangzhuo Xing$^4$, Zhixiang Shi$^4$ and Tsuyoshi Tamegai$^5$ }

\affiliation{%
$^1$Institute for Solid State Physics (ISSP), The University of Tokyo, Kashiwa, Chiba 277-8581, Japan\\
$^2$Department of Physics, Ritsumeikan University, Kusatsu, Shiga 525-8577, Japan\\
$^3$National Laboratory of Solid State Microstructures and Department of Physics, Nanjing University, Nanjing 210093, China\\
$^4$School of Physics and Key Laboratory of MEMS of the Ministry of Education, Southeast University, Nanjing 211189, China\\
$^5$Department of Applied Physics, The University of Tokyo, Bunkyo-ku, Tokyo 113-8656, Japan\\
$^6$Department of Physics and Mathematics, Aoyama Gakuin University, Sagamihara 252-5258, Japan}


\begin{abstract}
In the electronic nematic state, an electronic system has a lower symmetry than the crystal structure of the same system. Electronic nematic states have been observed in various unconventional superconductors such as cuprate- and iron-based, heavy-fermion, and topological superconductors. The relation between nematicity and superconductivity is a major unsolved problem in condensed matter physics. By angle-resolved specific heat measurements, we report bulk quasi-particle evidence of nematicity in the topological superconductor Sr$_x$Bi$_2$Se$_3$. The specific heat exhibited a clear 2-fold symmetry despite the 3-fold symmetric lattice. Most importantly, the 2-fold symmetry appeared in the normal state above the superconducting transition temperature. This is explained by the angle-dependent Zeeman effect due to the anisotropic density of states in the nematic phase. Such results highlight the interrelation between nematicity and unconventional superconductivity.

\end{abstract}%

\maketitle

Electronic nematic phases, which break the rotational symmetry, have been proposed in many unconventional superconductors, such as high-$T_c$ cuprates \cite{YBCOnematicNature}, iron-based superconductors \cite{ChuScience2010,KasaharaBa122nameticnature}, and heavy-fermion materials \cite{OkazakiURuSinematicScience,RonningheavyFermionNature}. These discoveries suggest a generic origin of the superconducting (SC) pairing mechanism in correlated electron systems, which is strongly related to nematicity \cite{IBSsNematicreviewNatPhys}. In the cuprate superconductor YBa$_2$Cu$_3$O$_{7-\delta}$, the nematic state is observed at the same temperature as the pseudogap state, which is well above the superconducting dome \cite{YBCOnematicNature}. In the phase diagram of iron-based superconductors, such as electron-doped BaFe$_2$As$_2$, the nametic state resides in the temperature region above the antiferromagnetic and superconducting phases \cite{KasaharaBa122nameticnature}. The nematic phases of heavy-fermions also appear at temperatures far above the superconducting transition temperature, $T_{\rm{c}}$ \cite{OkazakiURuSinematicScience,RonningheavyFermionNature}. The nematic phases in correlated systems always seem to precede the superconductivity. This finding is crucial to elucidating the role of nematicity in establishing superconductivity (supporting, competing, or accidentally co-existing).

Recently, the nematic phase has also been observed in the topological superconductors M$_x$Bi$_2$Se$_3$ (M=Cu, Sr, Nb). Based on their strong spin-orbital couping (SOC) and multi-orbital effect \cite{FuLiangPRL,FuLiangPRB}, these superconductors are theoretically predicted to possess a fully gapped order parameter but odd-parity, i.e. two-fold symmetry of the SC gap. The nematic state was first demonstrated by nuclear magnetic resonance (NMR) measurements of Cu$_x$Bi$_2$Se$_3$: the spin susceptibility exhibited a 2-fold symmetry in the SC state under a rotating magnetic field in the hexagonal plane \cite{MatanoNMRNatPhys}. Subsequently,  various angle-resolved techniques have confirmed the nematic state in all M$_x$Bi$_2$Se$_3$ compounds. Such techniques involve upper critical fields \cite{PanSciRepsSrBiSeAngleHc2,DuScienceChinaSrBiSe,ShenJunying_NPJQM_NbBiSeAngleHc2,SmylieSciRep}, specific heat \cite{YonezawaARSHCuBiSe}, torque effects \cite{AsabaNbBiSePRX}, magnetization \cite{SmylieSciRep}, and scanning tunneling microscopy (STM) \cite{TaoRanPhysRevX.8.041024}. Interestingly, most of these reports found a spontaneous emergence of 2-fold symmetry with superconductivity. As a typical example, the angle-dependent specific heat measurements of Cu$_x$Bi$_2$Se$_3$, revealed a 2-fold quasi-particle (QP) oscillation, but only in the SC state \cite{YonezawaARSHCuBiSe}. This kind of nematic phase in topological superconductors, which is called \textit{nematic superconductivity}, obviously differs from the nematicity in other systems, such as cuprates, iron-based, and heavy-fermion superconductors. This difference is crucial for understanding the relation between nematicity and superconductivity.                         

However, the nature of nematic superconductivity remains debate. Angle dependent NMR and specific heat results yield different directions of the symmetric axis of Cu$_x$Bi$_2$Se$_3$ \cite{MatanoNMRNatPhys,YonezawaARSHCuBiSe}. Moreover, as revealed in recent angle-dependent magnetoresistance measurements, small single crystals of Sr$_x$Bi$_2$Se$_3$ also exhibit a 2-fold symmetry above $T_{\rm{c}}$ \cite{KuntsevichNJP}. However, the transport result is sensitive to the current path, which may be largely affected by the multi-domain effect as well as by interface effects between the domains \cite{KuntsevichNJP}. Therefore, to confirm whether nematic superconductivity is common to all topological superconductors, we require bulk evidence from QPs.

\begin{figure*}\center
	\includegraphics[width=16cm]{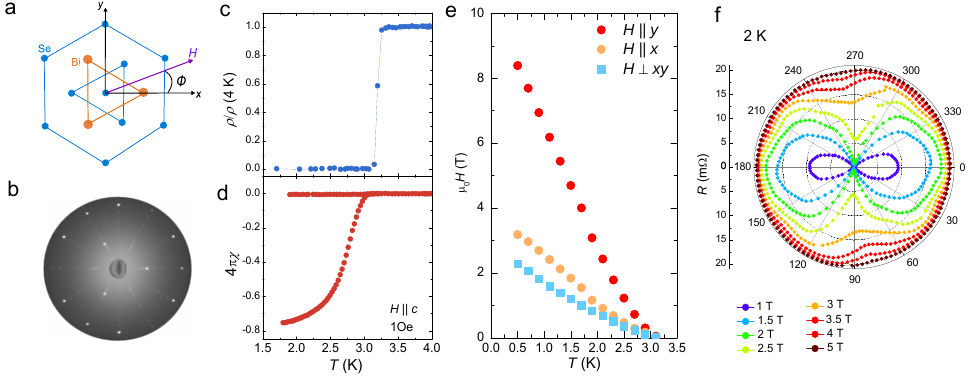}\\
	\caption{(a) Crystal structure of Sr$_x$Bi$_2$Se$_3$ in the hexagonal plane. The orange and blue spheres represent the elements Bi and Se. The $x$- and $y$-axes and the azimuthal angle $\phi$ are defined in the main text. (b) Laue diffraction pattern of the analyzed crystal. (c) Temperature dependence of the in-plane resistivity measured at zero field. (d) Temperature dependence of the magnetic susceptibility $\chi$ measured under 1 Oe field perpendicular to the $xy$-plane. (e) Upper critical fields for $H\parallel x$-axis, $H\parallel y$-axis and $H\perp xy$-plane, obtained from the resistive transition under fields. (f) Polar plot of resistance versus angle $\phi$ at 2 K under various magnetic fields (1 $\sim$ 5 T).}\label{}
\end{figure*}

In this Letter, we investigate the nematic state of Sr$_x$Bi$_2$Se$_3$ through angle-resolved specific heat (ARSH)-measurements. The SC volume \cite{LiuSrBiSeJACS} and in-plane anisotropy \cite{PanSciRepsSrBiSeAngleHc2} are reportedly larger in Sr$_x$Bi$_2$Se$_3$ than in Cu$_x$Bi$_2$Se$_3$. These properties are advantageous for studying the nematic phase. The specific heat of Sr$_x$Bi$_2$Se$_3$ exhibited a clear two-fold symmetry in both the superconducting and normal states, unambiguously providing QP evidence of the nematic state above $T_{\rm{c}}$.     

Single crystals of superconducting Sr$_x$Bi$_2$Se$_3$ (nominal composition $x$ = 0.1) were grown by the flux method \cite{LiuSrBiSeJACS}. The crystal structure of the sample was investigated by a Laue X-ray imaging system (Photonic Science Ltd). The magnetization was measured by a commercial SQUID magnetometer (MPMS-XL5, Quantum Design). The resistivities under magnetic fields (up to 9 T) were measured by the four-probe method in a physical property measurement system (PPMS, Quantum Design). The field-orientation dependence of the specific heat was measured in an 8 T split-pair superconducting magnet with a $^3$He refrigerator. The refrigerator can be continuously rotated by a motor on top of the dewar with an angular resolution better than 0.01$^\circ$. The calibration and validity of the measurement system are explained in the Supplemental Material S1 \cite{supplement}.

Sr$_x$Bi$_2$Se$_3$ consists of triangular-lattice layers of Bi and Se intercalated with Sr. Figure. 1(a) shows the crystal structure looking down the hexagonal plane. The structure is obviously hexagonal with 3-fold symmetry. The crystal structure of the analyzed sample was confirmed by the Laue diffraction pattern (see fig. 1(b)). The $x$-axis is defined as the Se-Bi bond direction in the hexagonal plane, and the $y$-axis is the in-plane direction perpendicular to the $x$-axis. Meanwhile, $\phi$ defines the azimuthal angle of the magnetic field with respect to the $x$-axis (Fig. 1(a)), and $\theta$ is the polar-angle of the field from the hexagonal plane. 

The superconducting transition temperature $T_{\rm{c}}$ of Sr$_x$Bi$_2$Se$_3$, obtained from the temperature dependences of resistivity (Fig. 1(c)) and susceptibility (Fig. 1(d)), was $\sim$ 3.1 K. Here, $T_{\rm{c}}$ defines the onset temperature of zero resistivity and deviation of the zero-field-cooling and field-cooling susceptibilities. Similar $T_{\rm{c}}$ were reported in previous studies \cite{LiuSrBiSeJACS,Shruti_SrBiSe_PhysRevB.92.020506,PanSciRepsSrBiSeAngleHc2,DuScienceChinaSrBiSe}. The shielding volume fraction approached 80\% at 1.8 K, confirming the bulk nature of the superconductivity. The upper critical fields for the magnetic fields in the directions parallel to ($H\parallel x$ and $y$) and perpendicular to ($H\perp xy$) the hexagonal plane were obtained from the resistive transitions under the respective fields, and are plotted as functions of temperature in Fig. 1(e). Besides the anisotropy in the in-plane and out-of-plane applied fields, the $H_{\rm{c2}}$ manifests obvious anisotropy within the hexagonal plane ($H_{\rm{c2}}^x$/$H_{\rm{c2}}^y$) $\sim$ 2.5 at low temperatures. Indeed, the angle dependent resistivity under a rotating field in the hexagonal plane manifests 2-fold symmetry (see Fig. 1(f)), which is similar to previous reports \cite{PanSciRepsSrBiSeAngleHc2,DuScienceChinaSrBiSe}.

\begin{figure*}\center
\includegraphics[width=17cm]{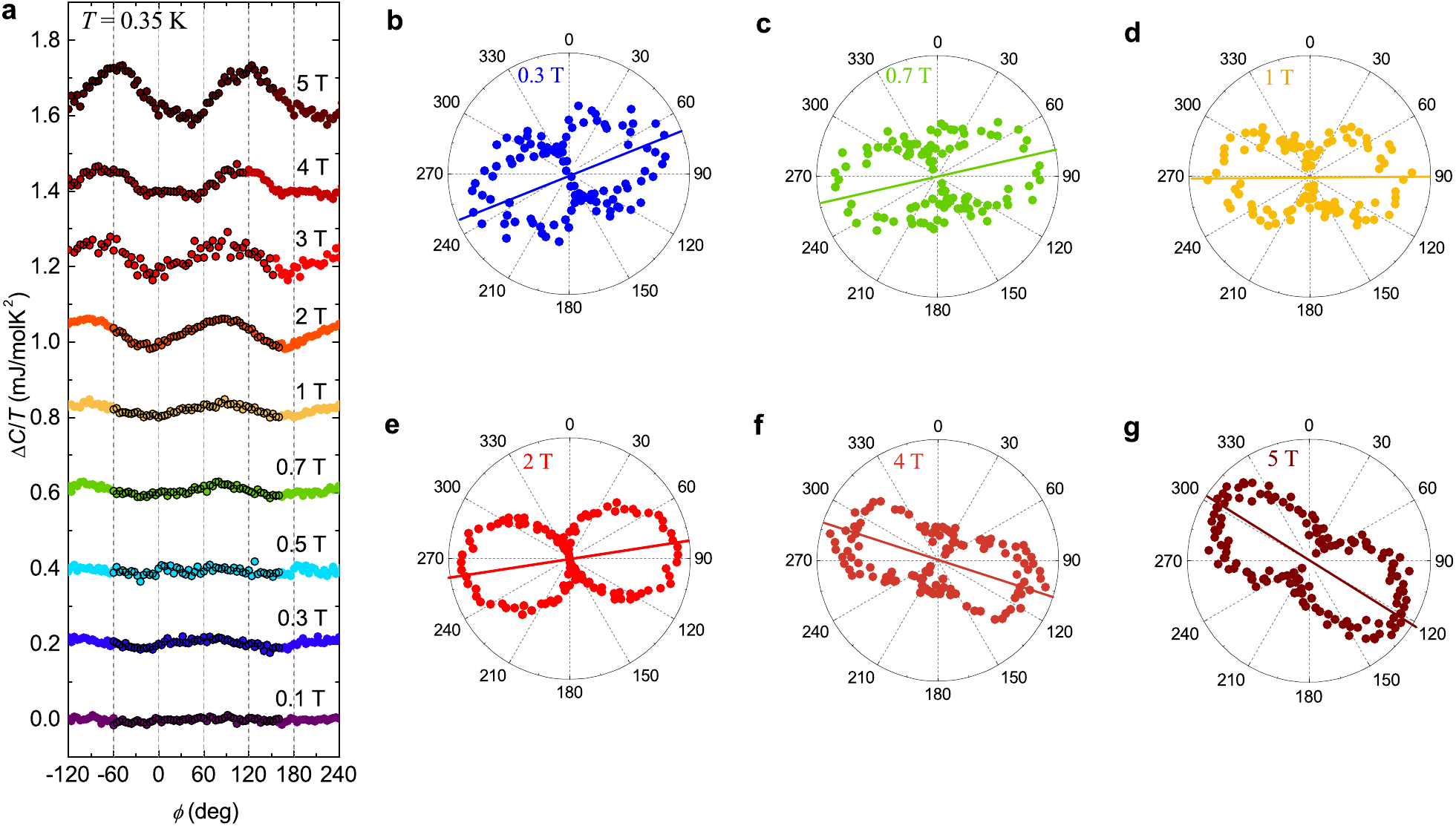}\\
\caption{(a) Azimuthal angle dependence of the specific heat $\Delta C(\phi)/T$ measured under various magnetic fields at 0.35 K. $\Delta C(\phi)/T$ is defined as $C(\phi)/T$-$C(0^\circ)/T$, and each subsequent curve is shifted vertically by 0.2 mJ/molK$^2$. Symbols with black outlines are measured data, and those without are mirrored points to show the symmetry. (b)-(g) Polar plots of $\Delta C(\phi)/T$ at 0.35 K under various magnetic fields: 0.3 T (b), 0.7 T (c), 1 T (d), 2 T (e), 4 T (f), and 5 T (g). The colored lines indicate the symmetry axes of the 2-fold oscillation.}\label{}
\end{figure*}

Figure 2(a) shows the azimuthal angle-resolved $\Delta C(\phi)/T$ values at 0.35 K under magnetic field ranging from 0.1 to 5 T. The $\Delta C(\phi)/T$ manifests a clear 2-fold symmetry, which is easily recognized in the polar plots (shown for selected fields in Figs. 2(b)-(g). Such 2-fold symmetry is consistent with that observed in the angle-dependent resistivity measurements (Fig. 1(f)). As the ARSH measurements probe the bulk signal of QPs, our results provide bulk evidence of nemacity in Sr$_x$Bi$_2$Se$_3$. Similar observations were reported in a previous study of Cu$_x$Bi$_2$Se$_3$ \cite{YonezawaARSHCuBiSe}. When the field exceeded 4 T, parts of $\Delta C(\phi)/T$ probed the normal state properties, which may explain the irregular behavior observed at 4 T. Unexpectedly, increasing the magnetic field caused a directional shift in the symmetry axis, as indicated by the lines in Figs. 2(b)-(g). Recent STM measurements of Cu$_x$Bi$_2$Se$_3$ revealed the same behavior \cite{TaoRanPhysRevX.8.041024}, possibly manifesting from a multi-domain effect, as discussed later.          

\begin{figure*}\center
\includegraphics[width=17cm]{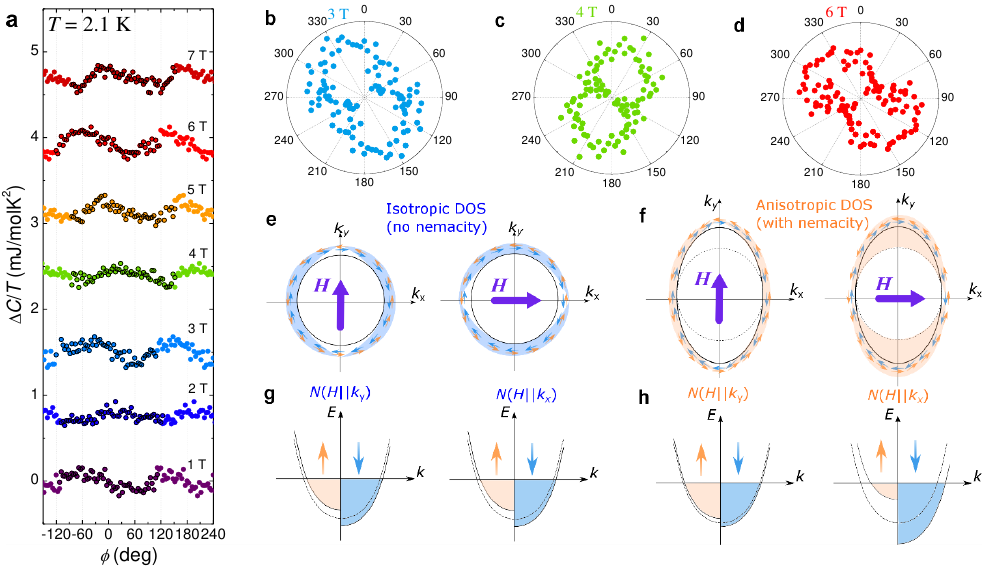}\\
\caption{(a) Azimuthal angle dependence of the specific heat $\Delta C(\phi)/T$ that is measured under various magnetic fields at 2.1 K. $\Delta C(\phi)/T$ is defined as $C(\phi)/T$-$C(0^\circ)/T$, and each subsequent curve is shifted vertically by 0.8 mJ/molK$^2$. Symbols with black outlines are measured data, and those without are mirrored points to show the symmetry. (b)-(d) Polar plots of $\Delta C(\phi)/T$ at 2.1 K under fields of 3 T, 4 T and 6 T, respectively. Sketches of the spin-helical structure in the situations of (e) isotropic density of states without nematicity, and (f) anisotropic density of states with nematicity. Paired spins are indicated by orange and blue arrows. The blue and orange patterns in (e) and (f) represent the spatial distributions of the magnitude of the Zeeman effect under fields along the $k_y$ and $k_x$ directions, respectively. Sketches of the band shift due to the Zeeman effect under fields along the $k_y$ and $k_x$ directions in the situations of (g) isotropic density of states without nematicity, and (h) anisotropic density of states with nematicity.}\label{}
\end{figure*}

Figure 3(a) shows the $\Delta C(\phi)/T$ values measured at 2.1 K under various magnetic fields (1 - 7 T). At 2.1 K, the in-plane $H_{\rm{c2}}$ under both $H\parallel x$ and $H\parallel y$ was below 3 T (see Fig. 1(e)). Thus, the $\Delta C(\phi)/T$ measured at $H \geq$ 3 T can probe the normal-state properties of Sr$_x$Bi$_2$Se$_3$. Astonishingly, $\Delta C(\phi)/T$ in the normal state also manifested 2-fold symmetry, as clarified in the polar plots of selected data (Figs. 3(b)-(d) under 3, 4, and 6 T, respectively). 2-fold symmetry in the normal state was also confirmed at $T$ = 3.5 K above $T_{\rm{c}}$ (see Supplemental Material S2 \cite{supplement}). To confirm that the observed 2-fold symmetry is the intrinsic property of the crystal, we carefully checked that our measurement-system contributes no oscillating background (Supplemental Material S1 \cite{supplement}). We also measured the polar-angle dependence of the specific heat $\Delta C(\theta)/T$ in the normal state. The $\Delta C(\theta)/T$ was $\theta$-independent, as presented in Supplemental Material S3 \cite{supplement}. Therefore, the in-plane two-fold symmetry cannot be attributed to an out-of-plane signal caused by misalignment of the sample setting. Besides, the Schottky anomaly is only observed at temperatures below 0.8 K (Supplemental Material S4 \cite{supplement}), which will not affect the observation of 2-fold symmetry in the normal state. The above observation unambiguously proves that the nematic state emerges from the normal state at temperatures above $T_{\rm{c}}$.       

To understand the observable normal-state nematicity in QPs measurements under a magnetic field, we should consider the special spin structure. Owing to the strong SOC, the spin direction is locked perpendicular to the momentum direction, forming the well-known spin-helical structure \cite{FuLiangPRL,JozwiakNatPhys}. In an isotropic density of states (DOS) without nemacity, the spins are homogeneously distributed as schematized in Fig. 3(e). As the specific heat is a bulk measurement, the probed QPs should be mainly contributed by the bulk band, in which the up and down spins are paired as represented by the paired orange and blue arrows in Fig. 3(e). In the surface state, the spins are polarized \cite{HsiehNature_BiSeSpin} but still form the helical structure; hence, they will not affect our following discussion.

In an isotropic DOS, magnetic field along the $k_y$ direction causes a strong Zeeman effect on the QPs with spins along the $k_x$ and -$k_x$ directions because the Zeeman effect is proportional to the magnitudes of the spins parallel to the field. On the contrary, a field along the $k_x$ direction induces a strong Zeeman effect on the QPs with spins along the $k_y$ and -$k_y$ directions. The Zeeman effect is represented by the blue regions in Fig. 3(e). Although the maximum and minimum positions change when the field rotation shifts from the $k_y$ to the $k_x$ directions, the total magnitude of the Zeeman effect (area of blue region in Fig. 3(e)) is unchanged because of the isotropic spin structure. Thus, the change of QPs due to the band shift under the Zeeman effect is independent of the in-plane direction of the field (see Fig. 3(g); the constructed band structure was based on the angle-resolved photoemission spectroscopy results \cite{SrBiSeARPESAPL}). On contrary, in case of anisotropic DOS with nemacity (Fig. 3(f)), more spins are arranged in the $\pm k_y$ directions in the top and bottom parts than in the $\pm k_x$ directions on the left and right sides, reflecting a distorted shape of the DOS. In this case, the Zeeman effect is much larger under $H \parallel k_x$ than under $H \parallel k_y$ (orange regions in Fig. 3(f)). Hence, the change in QPs due to the band shift under the Zeeman effect depends on the direction of the applied field (see Fig. 3(h)). When the field rotates in the $k_xk_y$-plane, the magnitude of the Zeeman effect manifests as a 2-fold symmetric oscillation. Under this 2-fold symmetric Zeeman effect, the nematicity in the normal state becomes observable in angle-resolved QPs measurements.

In a normal metal like Cu, the Zeeman effect is too small to be detected because the energy from applied field (1 T : $\sim$ 5.8$\times$10$^{-5}$ eV) is much smaller than the Fermi energy $\varepsilon_F$ (Cu: $\sim$ 7 eV). However, the doped Bi$_2$Se$_3$ exhibits a low carrier density $n$ ($\sim$ 10$^{20}$ cm$^{-3}$ for Cu$_x$Bi$_2$Se$_3$ \cite{HorPhysRevLett.104.057001}, and $\sim$ 10$^{19}$ cm$^{-3}$ for Sr$_x$Bi$_2$Se$_3$ \cite{LiuSrBiSeJACS}), which is much smaller than that of the normal metal (10$^{23}$ cm$^{-3}$ for Cu). Since the $\varepsilon_F$ is proportional to $n^{2/3}$, $\varepsilon_F$ of Sr$_x$Bi$_2$Se$_3$ can be simply estimated as $\varepsilon_F$(Sr$_x$Bi$_2$Se$_3$) = [$n$(Sr$_x$Bi$_2$Se$_3$)/$n$(Cu)]$^{2/3}$ $\times$ $\varepsilon_F$(Cu) $\sim$ 0.015 eV. Thus, several-Tesla magnetic field is comparable to a few percentage of $\varepsilon_F$, which is large enough to be measured.

Note that the above explanation of why the normal-state nematicity can be observed through the Zeeman effect is very generic in a strong SOC system. The SC jump in specific heat ($\Delta C$/$T_{\rm{c}}$) is very small ($\sim$ 1.1 mJ/mol K$^2$; see Supplemental Material S5 \cite{supplement}). The small jump is also reported previously as 1.4 mJ/mol K$^2$ in Ref. \cite{YonezawaARSHCuBiSe} and 0.28 mJ/mol K$^2$ in Ref. \cite{KristinSrBiSeARSHPhysRevB.98.184509}. Conversely, the magnitude of the oscillation in $\Delta C(\phi)/T$ is as large as $\sim$ 0.1 mJ/mol K$^2$ (Fig. 2(a)), approaching 10\% of the $\Delta C$/$T_{\rm{c}}$, much larger than in other superconductors such as heavy-fermions \cite{SakakibaraReview}, and iron-based superconductors \cite{YueSunPRBFeSeARSH}. Thus, the 2-fold symmetries observed in the SC and normal state of Sr$_x$Bi$_2$Se$_3$ may share a common origin. According to the theoretical calculation by M. Hecker and J. Schmalian \cite{HeckerNPJQM}, the nematic phase in M$_x$Bi$_2$Se$_3$ may originate from the SC fluctuation, which begins above $T_{\rm{c}}$, but its strength decreases quickly in the normal state. Thus, the nematic phase exists at temperatures only slightly higher than $T_{\rm{c}}$. As stated above, we did observe the nematic state at 3.5 K above $T_{\rm{c}}$. We also observed no symmetric oscillation at 5 K (See Fig. S3(e) \cite{supplement}), which is consistent to the theory.              

In the normal state, the 2-fold symmetry was observed only in the ARSH measurements, and was absent in the transport measurements such as the anisotropy measurement of $H_{\rm{c2}}$ in our crystal (Fig. 1(f)) and in most other reports \cite{PanSciRepsSrBiSeAngleHc2,DuScienceChinaSrBiSe,ShenJunying_NPJQM_NbBiSeAngleHc2}. It may originate from the multi-domain effect as reported in \cite{KuntsevichNJP}. In the normal state, the current path will be largely affected by the domains and the interface effect between the domains (for instance, the $T_{\rm{c}}$ is higher at the interfaces than the bulk) \cite{KuntsevichNJP}. Therefore, the current direction becomes randomly distributed and the 2-fold symmetry cancels out. The strongest evidence of 2-fold symmetry in the normal state appears in the transport measurements of a very small crystal with a single domain \cite{KuntsevichNJP}. Considering the multi-domain effect, the irregular change of the symmetry axis as well as the magnitude of the oscillation by magnetic fields can be also explained (see Supplemental Material S6 \cite{supplement}). As the symmetry axis also changes in the $H_{\rm{c2}}$ \cite{DuScienceChinaSrBiSe} and STM measurements \cite{TaoRanPhysRevX.8.041024}, it appears to be a common feature of M$_x$Bi$_2$Se$_3$, and must be investigated in future efforts.    

Finally, we emphasize that the nematicity above $T_{\rm{c}}$ observed in Sr$_x$Bi$_2$Se$_3$ differs from that in the similar compound Cu$_x$Bi$_2$Se$_3$, which exhibits nematicity only in the SC state \cite{YonezawaARSHCuBiSe}. In fact, Sr$_x$Bi$_2$Se$_3$ shows a much larger in-plane $H_{\rm{c2}}$ anisotropy ($\sim$ 2.5; see Fig. 1(e) and \cite{PanSciRepsSrBiSeAngleHc2}) than Cu$_x$Bi$_2$Se$_3$ ($\sim$ 1.3; \cite{YonezawaARSHCuBiSe}). This implies that the DOS in Cu$_x$Bi$_2$Se$_3$ is nearly isotropic (consistent to the isotropic Knight shift in the normal state of Ref. \cite{MatanoNMRNatPhys}), and that the anisotropy is too weak to be distinguished in QPs measurements. It should be noted that the 2-fold symmetry at the normal state of Sr$_x$Bi$_2$Se$_3$ was not observed in the magnetization and specific heat measurements from another group \cite{SmylieSciRep,KristinSrBiSeARSHPhysRevB.98.184509}, which may be due to the multi-domain effect or the weakened superconducting fluctuation because the specific heat jump at $T_{\rm{c}}$ in Ref. \cite{KristinSrBiSeARSHPhysRevB.98.184509} is only 0.28 mJ/mol K$^2$, much smaller than that of 1.1 mJ/mol K$^2$ in this work.

The authors would like to thank Y. Qiu and S. Nakatsuji from ISSP of the University of Tokyo for help with the Laue diffraction measurements. They would also like to thank J\"{o}rg Schmalian, Liang Fu, Peng Zhang, Guoqing Zheng, and Yuji Matsuda for the helpful discussions. The present work was partly supported by ``J-Physics'' (15H05883 and 18H04306), KAKENHI (19K14661, 17H01141, 18H01161, and JP17K05553) from JSPS, and NSFC of China (11822405, 11674157, 11674054, and  11611140101).

\bibliography{nematicSrBiSereferences}

\begin{thebibliography}{28}%
\makeatletter
\providecommand \@ifxundefined [1]{%
 \@ifx{#1\undefined}
}%
\providecommand \@ifnum [1]{%
 \ifnum #1\expandafter \@firstoftwo
 \else \expandafter \@secondoftwo
 \fi
}%
\providecommand \@ifx [1]{%
 \ifx #1\expandafter \@firstoftwo
 \else \expandafter \@secondoftwo
 \fi
}%
\providecommand \natexlab [1]{#1}%
\providecommand \enquote  [1]{``#1''}%
\providecommand \bibnamefont  [1]{#1}%
\providecommand \bibfnamefont [1]{#1}%
\providecommand \citenamefont [1]{#1}%
\providecommand \href@noop [0]{\@secondoftwo}%
\providecommand \href [0]{\begingroup \@sanitize@url \@href}%
\providecommand \@href[1]{\@@startlink{#1}\@@href}%
\providecommand \@@href[1]{\endgroup#1\@@endlink}%
\providecommand \@sanitize@url [0]{\catcode `\\12\catcode `\$12\catcode
  `\&12\catcode `\#12\catcode `\^12\catcode `\_12\catcode `\%12\relax}%
\providecommand \@@startlink[1]{}%
\providecommand \@@endlink[0]{}%
\providecommand \url  [0]{\begingroup\@sanitize@url \@url }%
\providecommand \@url [1]{\endgroup\@href {#1}{\urlprefix }}%
\providecommand \urlprefix  [0]{URL }%
\providecommand \Eprint [0]{\href }%
\providecommand \doibase [0]{http://dx.doi.org/}%
\providecommand \selectlanguage [0]{\@gobble}%
\providecommand \bibinfo  [0]{\@secondoftwo}%
\providecommand \bibfield  [0]{\@secondoftwo}%
\providecommand \translation [1]{[#1]}%
\providecommand \BibitemOpen [0]{}%
\providecommand \bibitemStop [0]{}%
\providecommand \bibitemNoStop [0]{.\EOS\space}%
\providecommand \EOS [0]{\spacefactor3000\relax}%
\providecommand \BibitemShut  [1]{\csname bibitem#1\endcsname}%
\let\auto@bib@innerbib\@empty
\bibitem [{\citenamefont {Daou}\ \emph {et~al.}(2010)\citenamefont {Daou},
  \citenamefont {Chang}, \citenamefont {LeBoeuf}, \citenamefont
  {Cyr-Choiniere}, \citenamefont {Laliberte}, \citenamefont {Doiron-Leyraud},
  \citenamefont {Ramshaw}, \citenamefont {Liang}, \citenamefont {Bonn},
  \citenamefont {Hardy},\ and\ \citenamefont {Taillefer}}]{YBCOnematicNature}%
  \BibitemOpen
  \bibfield  {author} {\bibinfo {author} {\bibfnamefont {R.}~\bibnamefont
  {Daou}}, \bibinfo {author} {\bibfnamefont {J.}~\bibnamefont {Chang}},
  \bibinfo {author} {\bibfnamefont {D.}~\bibnamefont {LeBoeuf}}, \bibinfo
  {author} {\bibfnamefont {O.}~\bibnamefont {Cyr-Choiniere}}, \bibinfo {author}
  {\bibfnamefont {F.}~\bibnamefont {Laliberte}}, \bibinfo {author}
  {\bibfnamefont {N.}~\bibnamefont {Doiron-Leyraud}}, \bibinfo {author}
  {\bibfnamefont {B.~J.}\ \bibnamefont {Ramshaw}}, \bibinfo {author}
  {\bibfnamefont {R.}~\bibnamefont {Liang}}, \bibinfo {author} {\bibfnamefont
  {D.~A.}\ \bibnamefont {Bonn}}, \bibinfo {author} {\bibfnamefont {W.~N.}\
  \bibnamefont {Hardy}}, \ and\ \bibinfo {author} {\bibfnamefont
  {L.}~\bibnamefont {Taillefer}},\ }\href@noop {} {\bibfield  {journal}
  {\bibinfo  {journal} {Nature}\ }\textbf {\bibinfo {volume} {463}},\ \bibinfo
  {pages} {519} (\bibinfo {year} {2010})}\BibitemShut {NoStop}%
\bibitem [{\citenamefont {Chu}\ \emph {et~al.}(2010)\citenamefont {Chu},
  \citenamefont {Analytis}, \citenamefont {De~Greve}, \citenamefont {McMahon},
  \citenamefont {Islam}, \citenamefont {Yamamoto},\ and\ \citenamefont
  {Fisher}}]{ChuScience2010}%
  \BibitemOpen
  \bibfield  {author} {\bibinfo {author} {\bibfnamefont {J.-H.}\ \bibnamefont
  {Chu}}, \bibinfo {author} {\bibfnamefont {J.~G.}\ \bibnamefont {Analytis}},
  \bibinfo {author} {\bibfnamefont {K.}~\bibnamefont {De~Greve}}, \bibinfo
  {author} {\bibfnamefont {P.~L.}\ \bibnamefont {McMahon}}, \bibinfo {author}
  {\bibfnamefont {Z.}~\bibnamefont {Islam}}, \bibinfo {author} {\bibfnamefont
  {Y.}~\bibnamefont {Yamamoto}}, \ and\ \bibinfo {author} {\bibfnamefont
  {I.~R.}\ \bibnamefont {Fisher}},\ }\href@noop {} {\bibfield  {journal}
  {\bibinfo  {journal} {Science}\ }\textbf {\bibinfo {volume} {329}},\ \bibinfo
  {pages} {824} (\bibinfo {year} {2010})}\BibitemShut {NoStop}%
\bibitem [{\citenamefont {Kasahara}\ \emph {et~al.}(2012)\citenamefont
  {Kasahara}, \citenamefont {Shi}, \citenamefont {Hashimoto}, \citenamefont
  {Tonegawa}, \citenamefont {Mizukami}, \citenamefont {Shibauchi},
  \citenamefont {Sugimoto}, \citenamefont {Fukuda}, \citenamefont {Terashima},
  \citenamefont {Nevidomskyy},\ and\ \citenamefont
  {Matsuda}}]{KasaharaBa122nameticnature}%
  \BibitemOpen
  \bibfield  {author} {\bibinfo {author} {\bibfnamefont {S.}~\bibnamefont
  {Kasahara}}, \bibinfo {author} {\bibfnamefont {H.~J.}\ \bibnamefont {Shi}},
  \bibinfo {author} {\bibfnamefont {K.}~\bibnamefont {Hashimoto}}, \bibinfo
  {author} {\bibfnamefont {S.}~\bibnamefont {Tonegawa}}, \bibinfo {author}
  {\bibfnamefont {Y.}~\bibnamefont {Mizukami}}, \bibinfo {author}
  {\bibfnamefont {T.}~\bibnamefont {Shibauchi}}, \bibinfo {author}
  {\bibfnamefont {K.}~\bibnamefont {Sugimoto}}, \bibinfo {author}
  {\bibfnamefont {T.}~\bibnamefont {Fukuda}}, \bibinfo {author} {\bibfnamefont
  {T.}~\bibnamefont {Terashima}}, \bibinfo {author} {\bibfnamefont {A.~H.}\
  \bibnamefont {Nevidomskyy}}, \ and\ \bibinfo {author} {\bibfnamefont
  {Y.}~\bibnamefont {Matsuda}},\ }\href@noop {} {\bibfield  {journal} {\bibinfo
   {journal} {Nature}\ }\textbf {\bibinfo {volume} {486}},\ \bibinfo {pages}
  {382} (\bibinfo {year} {2012})}\BibitemShut {NoStop}%
\bibitem [{\citenamefont {Okazaki}\ \emph {et~al.}(2011)\citenamefont
  {Okazaki}, \citenamefont {Shibauchi}, \citenamefont {Shi}, \citenamefont
  {Haga}, \citenamefont {Matsuda}, \citenamefont {Yamamoto}, \citenamefont
  {Onuki}, \citenamefont {Ikeda},\ and\ \citenamefont
  {Matsuda}}]{OkazakiURuSinematicScience}%
  \BibitemOpen
  \bibfield  {author} {\bibinfo {author} {\bibfnamefont {R.}~\bibnamefont
  {Okazaki}}, \bibinfo {author} {\bibfnamefont {T.}~\bibnamefont {Shibauchi}},
  \bibinfo {author} {\bibfnamefont {H.~J.}\ \bibnamefont {Shi}}, \bibinfo
  {author} {\bibfnamefont {Y.}~\bibnamefont {Haga}}, \bibinfo {author}
  {\bibfnamefont {T.~D.}\ \bibnamefont {Matsuda}}, \bibinfo {author}
  {\bibfnamefont {E.}~\bibnamefont {Yamamoto}}, \bibinfo {author}
  {\bibfnamefont {Y.}~\bibnamefont {Onuki}}, \bibinfo {author} {\bibfnamefont
  {H.}~\bibnamefont {Ikeda}}, \ and\ \bibinfo {author} {\bibfnamefont
  {Y.}~\bibnamefont {Matsuda}},\ }\href@noop {} {\bibfield  {journal} {\bibinfo
   {journal} {Science}\ }\textbf {\bibinfo {volume} {331}},\ \bibinfo {pages}
  {439} (\bibinfo {year} {2011})}\BibitemShut {NoStop}%
\bibitem [{\citenamefont {Ronning}\ \emph {et~al.}(2017)\citenamefont
  {Ronning}, \citenamefont {Helm}, \citenamefont {Shirer}, \citenamefont
  {Bachmann}, \citenamefont {Balicas}, \citenamefont {Chan}, \citenamefont
  {Ramshaw}, \citenamefont {McDonald}, \citenamefont {Balakirev}, \citenamefont
  {Jaime}, \citenamefont {Bauer},\ and\ \citenamefont
  {Moll}}]{RonningheavyFermionNature}%
  \BibitemOpen
  \bibfield  {author} {\bibinfo {author} {\bibfnamefont {F.}~\bibnamefont
  {Ronning}}, \bibinfo {author} {\bibfnamefont {T.}~\bibnamefont {Helm}},
  \bibinfo {author} {\bibfnamefont {K.~R.}\ \bibnamefont {Shirer}}, \bibinfo
  {author} {\bibfnamefont {M.~D.}\ \bibnamefont {Bachmann}}, \bibinfo {author}
  {\bibfnamefont {L.}~\bibnamefont {Balicas}}, \bibinfo {author} {\bibfnamefont
  {M.~K.}\ \bibnamefont {Chan}}, \bibinfo {author} {\bibfnamefont {B.~J.}\
  \bibnamefont {Ramshaw}}, \bibinfo {author} {\bibfnamefont {R.~D.}\
  \bibnamefont {McDonald}}, \bibinfo {author} {\bibfnamefont {F.~F.}\
  \bibnamefont {Balakirev}}, \bibinfo {author} {\bibfnamefont {M.}~\bibnamefont
  {Jaime}}, \bibinfo {author} {\bibfnamefont {E.~D.}\ \bibnamefont {Bauer}}, \
  and\ \bibinfo {author} {\bibfnamefont {P.~J.~W.}\ \bibnamefont {Moll}},\
  }\href@noop {} {\bibfield  {journal} {\bibinfo  {journal} {Nature}\ }\textbf
  {\bibinfo {volume} {548}},\ \bibinfo {pages} {313} (\bibinfo {year}
  {2017})}\BibitemShut {NoStop}%
\bibitem [{\citenamefont {Fernandes}\ \emph {et~al.}(2014)\citenamefont
  {Fernandes}, \citenamefont {Chubukov},\ and\ \citenamefont
  {Schmalian}}]{IBSsNematicreviewNatPhys}%
  \BibitemOpen
  \bibfield  {author} {\bibinfo {author} {\bibfnamefont {R.~M.}\ \bibnamefont
  {Fernandes}}, \bibinfo {author} {\bibfnamefont {A.~V.}\ \bibnamefont
  {Chubukov}}, \ and\ \bibinfo {author} {\bibfnamefont {J.}~\bibnamefont
  {Schmalian}},\ }\href@noop {} {\bibfield  {journal} {\bibinfo  {journal}
  {Nat. Phys.}\ }\textbf {\bibinfo {volume} {10}},\ \bibinfo {pages} {97}
  (\bibinfo {year} {2014})}\BibitemShut {NoStop}%
\bibitem [{\citenamefont {Fu}\ and\ \citenamefont {Berg}(2010)}]{FuLiangPRL}%
  \BibitemOpen
  \bibfield  {author} {\bibinfo {author} {\bibfnamefont {L.}~\bibnamefont
  {Fu}}\ and\ \bibinfo {author} {\bibfnamefont {E.}~\bibnamefont {Berg}},\
  }\href@noop {} {\bibfield  {journal} {\bibinfo  {journal} {Phys. Rev. Lett.}\
  }\textbf {\bibinfo {volume} {105}},\ \bibinfo {pages} {097001} (\bibinfo
  {year} {2010})}\BibitemShut {NoStop}%
\bibitem [{\citenamefont {Fu}(2014)}]{FuLiangPRB}%
  \BibitemOpen
  \bibfield  {author} {\bibinfo {author} {\bibfnamefont {L.}~\bibnamefont
  {Fu}},\ }\href@noop {} {\bibfield  {journal} {\bibinfo  {journal} {Phys. Rev.
  B}\ }\textbf {\bibinfo {volume} {90}},\ \bibinfo {pages} {100509} (\bibinfo
  {year} {2014})}\BibitemShut {NoStop}%
\bibitem [{\citenamefont {Matano}\ \emph {et~al.}(2016)\citenamefont {Matano},
  \citenamefont {Kriener}, \citenamefont {Segawa}, \citenamefont {Ando},\ and\
  \citenamefont {Zheng}}]{MatanoNMRNatPhys}%
  \BibitemOpen
  \bibfield  {author} {\bibinfo {author} {\bibfnamefont {K.}~\bibnamefont
  {Matano}}, \bibinfo {author} {\bibfnamefont {M.}~\bibnamefont {Kriener}},
  \bibinfo {author} {\bibfnamefont {K.}~\bibnamefont {Segawa}}, \bibinfo
  {author} {\bibfnamefont {Y.}~\bibnamefont {Ando}}, \ and\ \bibinfo {author}
  {\bibfnamefont {G.-q.}\ \bibnamefont {Zheng}},\ }\href@noop {} {\bibfield
  {journal} {\bibinfo  {journal} {Nat. Phys.}\ }\textbf {\bibinfo {volume}
  {12}},\ \bibinfo {pages} {852} (\bibinfo {year} {2016})}\BibitemShut
  {NoStop}%
\bibitem [{\citenamefont {Pan}\ \emph {et~al.}(2016)\citenamefont {Pan},
  \citenamefont {Nikitin}, \citenamefont {Araizi}, \citenamefont {Huang},
  \citenamefont {Matsushita}, \citenamefont {Naka},\ and\ \citenamefont
  {de~Visser}}]{PanSciRepsSrBiSeAngleHc2}%
  \BibitemOpen
  \bibfield  {author} {\bibinfo {author} {\bibfnamefont {Y.}~\bibnamefont
  {Pan}}, \bibinfo {author} {\bibfnamefont {A.~M.}\ \bibnamefont {Nikitin}},
  \bibinfo {author} {\bibfnamefont {G.~K.}\ \bibnamefont {Araizi}}, \bibinfo
  {author} {\bibfnamefont {Y.~K.}\ \bibnamefont {Huang}}, \bibinfo {author}
  {\bibfnamefont {Y.}~\bibnamefont {Matsushita}}, \bibinfo {author}
  {\bibfnamefont {T.}~\bibnamefont {Naka}}, \ and\ \bibinfo {author}
  {\bibfnamefont {A.}~\bibnamefont {de~Visser}},\ }\href@noop {} {\bibfield
  {journal} {\bibinfo  {journal} {Sci. Rep.}\ }\textbf {\bibinfo {volume}
  {6}},\ \bibinfo {pages} {28632} (\bibinfo {year} {2016})}\BibitemShut
  {NoStop}%
\bibitem [{\citenamefont {Du}\ \emph {et~al.}(2017)\citenamefont {Du},
  \citenamefont {Li}, \citenamefont {Schneeloch}, \citenamefont {Zhong},
  \citenamefont {Gu}, \citenamefont {Yang}, \citenamefont {Lin},\ and\
  \citenamefont {Wen}}]{DuScienceChinaSrBiSe}%
  \BibitemOpen
  \bibfield  {author} {\bibinfo {author} {\bibfnamefont {G.}~\bibnamefont
  {Du}}, \bibinfo {author} {\bibfnamefont {Y.}~\bibnamefont {Li}}, \bibinfo
  {author} {\bibfnamefont {J.}~\bibnamefont {Schneeloch}}, \bibinfo {author}
  {\bibfnamefont {R.~D.}\ \bibnamefont {Zhong}}, \bibinfo {author}
  {\bibfnamefont {G.}~\bibnamefont {Gu}}, \bibinfo {author} {\bibfnamefont
  {H.}~\bibnamefont {Yang}}, \bibinfo {author} {\bibfnamefont {H.}~\bibnamefont
  {Lin}}, \ and\ \bibinfo {author} {\bibfnamefont {H.-H.}\ \bibnamefont
  {Wen}},\ }\href@noop {} {\bibfield  {journal} {\bibinfo  {journal} {Sci.
  China Phys. Mech. Astron.}\ }\textbf {\bibinfo {volume} {60}},\ \bibinfo
  {pages} {037411} (\bibinfo {year} {2017})}\BibitemShut {NoStop}%
\bibitem [{\citenamefont {Shen}\ \emph {et~al.}(2017)\citenamefont {Shen},
  \citenamefont {He}, \citenamefont {Yuan}, \citenamefont {Huang},
  \citenamefont {Cho}, \citenamefont {Lee}, \citenamefont {Hor}, \citenamefont
  {Law},\ and\ \citenamefont {Lortz}}]{ShenJunying_NPJQM_NbBiSeAngleHc2}%
  \BibitemOpen
  \bibfield  {author} {\bibinfo {author} {\bibfnamefont {J.}~\bibnamefont
  {Shen}}, \bibinfo {author} {\bibfnamefont {W.-Y.}\ \bibnamefont {He}},
  \bibinfo {author} {\bibfnamefont {N.~F.~Q.}\ \bibnamefont {Yuan}}, \bibinfo
  {author} {\bibfnamefont {Z.}~\bibnamefont {Huang}}, \bibinfo {author}
  {\bibfnamefont {C.-w.}\ \bibnamefont {Cho}}, \bibinfo {author} {\bibfnamefont
  {S.~H.}\ \bibnamefont {Lee}}, \bibinfo {author} {\bibfnamefont {Y.~S.}\
  \bibnamefont {Hor}}, \bibinfo {author} {\bibfnamefont {K.~T.}\ \bibnamefont
  {Law}}, \ and\ \bibinfo {author} {\bibfnamefont {R.}~\bibnamefont {Lortz}},\
  }\href@noop {} {\bibfield  {journal} {\bibinfo  {journal} {npj Quantum
  Materials}\ }\textbf {\bibinfo {volume} {2}},\ \bibinfo {pages} {59}
  (\bibinfo {year} {2017})}\BibitemShut {NoStop}%
\bibitem [{\citenamefont {Smylie}\ \emph {et~al.}(2018)\citenamefont {Smylie},
  \citenamefont {Willa}, \citenamefont {Claus}, \citenamefont {Koshelev},
  \citenamefont {Song}, \citenamefont {Kwok}, \citenamefont {Islam},
  \citenamefont {Gu}, \citenamefont {Schneeloch}, \citenamefont {Zhong},\ and\
  \citenamefont {Welp}}]{SmylieSciRep}%
  \BibitemOpen
  \bibfield  {author} {\bibinfo {author} {\bibfnamefont {M.~P.}\ \bibnamefont
  {Smylie}}, \bibinfo {author} {\bibfnamefont {K.}~\bibnamefont {Willa}},
  \bibinfo {author} {\bibfnamefont {H.}~\bibnamefont {Claus}}, \bibinfo
  {author} {\bibfnamefont {A.~E.}\ \bibnamefont {Koshelev}}, \bibinfo {author}
  {\bibfnamefont {K.~W.}\ \bibnamefont {Song}}, \bibinfo {author}
  {\bibfnamefont {W.~K.}\ \bibnamefont {Kwok}}, \bibinfo {author}
  {\bibfnamefont {Z.}~\bibnamefont {Islam}}, \bibinfo {author} {\bibfnamefont
  {G.~D.}\ \bibnamefont {Gu}}, \bibinfo {author} {\bibfnamefont {J.~A.}\
  \bibnamefont {Schneeloch}}, \bibinfo {author} {\bibfnamefont {R.~D.}\
  \bibnamefont {Zhong}}, \ and\ \bibinfo {author} {\bibfnamefont
  {U.}~\bibnamefont {Welp}},\ }\href@noop {} {\bibfield  {journal} {\bibinfo
  {journal} {Sci. Rep.}\ }\textbf {\bibinfo {volume} {8}},\ \bibinfo {pages}
  {7666} (\bibinfo {year} {2018})}\BibitemShut {NoStop}%
\bibitem [{\citenamefont {Yonezawa}\ \emph {et~al.}(2016)\citenamefont
  {Yonezawa}, \citenamefont {Tajiri}, \citenamefont {Nakata}, \citenamefont
  {Nagai}, \citenamefont {Wang}, \citenamefont {Segawa}, \citenamefont {Ando},\
  and\ \citenamefont {Maeno}}]{YonezawaARSHCuBiSe}%
  \BibitemOpen
  \bibfield  {author} {\bibinfo {author} {\bibfnamefont {S.}~\bibnamefont
  {Yonezawa}}, \bibinfo {author} {\bibfnamefont {K.}~\bibnamefont {Tajiri}},
  \bibinfo {author} {\bibfnamefont {S.}~\bibnamefont {Nakata}}, \bibinfo
  {author} {\bibfnamefont {Y.}~\bibnamefont {Nagai}}, \bibinfo {author}
  {\bibfnamefont {Z.}~\bibnamefont {Wang}}, \bibinfo {author} {\bibfnamefont
  {K.}~\bibnamefont {Segawa}}, \bibinfo {author} {\bibfnamefont
  {Y.}~\bibnamefont {Ando}}, \ and\ \bibinfo {author} {\bibfnamefont
  {Y.}~\bibnamefont {Maeno}},\ }\href@noop {} {\bibfield  {journal} {\bibinfo
  {journal} {Nat. Phys.}\ }\textbf {\bibinfo {volume} {13}},\ \bibinfo {pages}
  {123} (\bibinfo {year} {2016})}\BibitemShut {NoStop}%
\bibitem [{\citenamefont {Asaba}\ \emph {et~al.}(2017)\citenamefont {Asaba},
  \citenamefont {Lawson}, \citenamefont {Tinsman}, \citenamefont {Chen},
  \citenamefont {Corbae}, \citenamefont {Li}, \citenamefont {Qiu},
  \citenamefont {Hor}, \citenamefont {Fu},\ and\ \citenamefont
  {Li}}]{AsabaNbBiSePRX}%
  \BibitemOpen
  \bibfield  {author} {\bibinfo {author} {\bibfnamefont {T.}~\bibnamefont
  {Asaba}}, \bibinfo {author} {\bibfnamefont {B.~J.}\ \bibnamefont {Lawson}},
  \bibinfo {author} {\bibfnamefont {C.}~\bibnamefont {Tinsman}}, \bibinfo
  {author} {\bibfnamefont {L.}~\bibnamefont {Chen}}, \bibinfo {author}
  {\bibfnamefont {P.}~\bibnamefont {Corbae}}, \bibinfo {author} {\bibfnamefont
  {G.}~\bibnamefont {Li}}, \bibinfo {author} {\bibfnamefont {Y.}~\bibnamefont
  {Qiu}}, \bibinfo {author} {\bibfnamefont {Y.~S.}\ \bibnamefont {Hor}},
  \bibinfo {author} {\bibfnamefont {L.}~\bibnamefont {Fu}}, \ and\ \bibinfo
  {author} {\bibfnamefont {L.}~\bibnamefont {Li}},\ }\href@noop {} {\bibfield
  {journal} {\bibinfo  {journal} {Phys. Rev. X}\ }\textbf {\bibinfo {volume}
  {7}},\ \bibinfo {pages} {011009} (\bibinfo {year} {2017})}\BibitemShut
  {NoStop}%
\bibitem [{\citenamefont {Tao}\ \emph {et~al.}(2018)\citenamefont {Tao},
  \citenamefont {Yan}, \citenamefont {Liu}, \citenamefont {Wang}, \citenamefont
  {Ando}, \citenamefont {Wang}, \citenamefont {Zhang},\ and\ \citenamefont
  {Feng}}]{TaoRanPhysRevX.8.041024}%
  \BibitemOpen
  \bibfield  {author} {\bibinfo {author} {\bibfnamefont {R.}~\bibnamefont
  {Tao}}, \bibinfo {author} {\bibfnamefont {Y.-J.}\ \bibnamefont {Yan}},
  \bibinfo {author} {\bibfnamefont {X.}~\bibnamefont {Liu}}, \bibinfo {author}
  {\bibfnamefont {Z.-W.}\ \bibnamefont {Wang}}, \bibinfo {author}
  {\bibfnamefont {Y.}~\bibnamefont {Ando}}, \bibinfo {author} {\bibfnamefont
  {Q.-H.}\ \bibnamefont {Wang}}, \bibinfo {author} {\bibfnamefont
  {T.}~\bibnamefont {Zhang}}, \ and\ \bibinfo {author} {\bibfnamefont {D.-L.}\
  \bibnamefont {Feng}},\ }\href {\doibase 10.1103/PhysRevX.8.041024} {\bibfield
   {journal} {\bibinfo  {journal} {Phys. Rev. X}\ }\textbf {\bibinfo {volume}
  {8}},\ \bibinfo {pages} {041024} (\bibinfo {year} {2018})}\BibitemShut
  {NoStop}%
\bibitem [{\citenamefont {Kuntsevich}\ \emph {et~al.}(2018)\citenamefont
  {Kuntsevich}, \citenamefont {Bryzgalov}, \citenamefont {Prudkoglyad},
  \citenamefont {Martovitskii}, \citenamefont {Selivanov},\ and\ \citenamefont
  {Chizhevskii}}]{KuntsevichNJP}%
  \BibitemOpen
  \bibfield  {author} {\bibinfo {author} {\bibfnamefont {A.~Y.}\ \bibnamefont
  {Kuntsevich}}, \bibinfo {author} {\bibfnamefont {M.~A.}\ \bibnamefont
  {Bryzgalov}}, \bibinfo {author} {\bibfnamefont {V.~A.}\ \bibnamefont
  {Prudkoglyad}}, \bibinfo {author} {\bibfnamefont {V.~P.}\ \bibnamefont
  {Martovitskii}}, \bibinfo {author} {\bibfnamefont {Y.~G.}\ \bibnamefont
  {Selivanov}}, \ and\ \bibinfo {author} {\bibfnamefont {E.~G.}\ \bibnamefont
  {Chizhevskii}},\ }\href {http://stacks.iop.org/1367-2630/20/i=10/a=103022}
  {\bibfield  {journal} {\bibinfo  {journal} {New J. Phys.}\ }\textbf {\bibinfo
  {volume} {20}},\ \bibinfo {pages} {103022} (\bibinfo {year}
  {2018})}\BibitemShut {NoStop}%
\bibitem [{\citenamefont {Liu}\ \emph {et~al.}(2015)\citenamefont {Liu},
  \citenamefont {Yao}, \citenamefont {Shao}, \citenamefont {Zuo}, \citenamefont
  {Pi}, \citenamefont {Tan}, \citenamefont {Zhang},\ and\ \citenamefont
  {Zhang}}]{LiuSrBiSeJACS}%
  \BibitemOpen
  \bibfield  {author} {\bibinfo {author} {\bibfnamefont {Z.}~\bibnamefont
  {Liu}}, \bibinfo {author} {\bibfnamefont {X.}~\bibnamefont {Yao}}, \bibinfo
  {author} {\bibfnamefont {J.}~\bibnamefont {Shao}}, \bibinfo {author}
  {\bibfnamefont {M.}~\bibnamefont {Zuo}}, \bibinfo {author} {\bibfnamefont
  {L.}~\bibnamefont {Pi}}, \bibinfo {author} {\bibfnamefont {S.}~\bibnamefont
  {Tan}}, \bibinfo {author} {\bibfnamefont {C.}~\bibnamefont {Zhang}}, \ and\
  \bibinfo {author} {\bibfnamefont {Y.}~\bibnamefont {Zhang}},\ }\href@noop {}
  {\bibfield  {journal} {\bibinfo  {journal} {J. Am. Chem. Soc.}\ }\textbf
  {\bibinfo {volume} {137}},\ \bibinfo {pages} {10512} (\bibinfo {year}
  {2015})}\BibitemShut {NoStop}%
\bibitem [{sup()}]{supplement}%
  \BibitemOpen
  \href@noop {} {\ }\bibinfo {note} {{See Supplemental Material at [] for the
  details about the experiment-system and calibration, azimuthal angle
  dependence of the specific heat $\Delta C(\phi)/T$ at 3.5 K and 5 K, the
  polar angle dependence of the specific heat $\Delta C(\theta)/T$, the
  Schottky anomaly at low temperature, the specific heat jump at $T_{\rm{c}}$,
  and the multi-domain effect.}}\BibitemShut {Stop}%
\bibitem [{\citenamefont {Shruti}\ \emph {et~al.}(2015)\citenamefont {Shruti},
  \citenamefont {Maurya}, \citenamefont {Neha}, \citenamefont {Srivastava},\
  and\ \citenamefont {Patnaik}}]{Shruti_SrBiSe_PhysRevB.92.020506}%
  \BibitemOpen
  \bibfield  {author} {\bibinfo {author} {\bibnamefont {Shruti}}, \bibinfo
  {author} {\bibfnamefont {V.~K.}\ \bibnamefont {Maurya}}, \bibinfo {author}
  {\bibfnamefont {P.}~\bibnamefont {Neha}}, \bibinfo {author} {\bibfnamefont
  {P.}~\bibnamefont {Srivastava}}, \ and\ \bibinfo {author} {\bibfnamefont
  {S.}~\bibnamefont {Patnaik}},\ }\href {\doibase 10.1103/PhysRevB.92.020506}
  {\bibfield  {journal} {\bibinfo  {journal} {Phys. Rev. B}\ }\textbf {\bibinfo
  {volume} {92}},\ \bibinfo {pages} {020506} (\bibinfo {year}
  {2015})}\BibitemShut {NoStop}%
\bibitem [{\citenamefont {Jozwiak}\ \emph {et~al.}(2013)\citenamefont
  {Jozwiak}, \citenamefont {Park}, \citenamefont {Gotlieb}, \citenamefont
  {Hwang}, \citenamefont {Lee}, \citenamefont {Louie}, \citenamefont
  {Denlinger}, \citenamefont {Rotundu}, \citenamefont {Birgeneau},
  \citenamefont {Hussain},\ and\ \citenamefont {Lanzara}}]{JozwiakNatPhys}%
  \BibitemOpen
  \bibfield  {author} {\bibinfo {author} {\bibfnamefont {C.}~\bibnamefont
  {Jozwiak}}, \bibinfo {author} {\bibfnamefont {C.-H.}\ \bibnamefont {Park}},
  \bibinfo {author} {\bibfnamefont {K.}~\bibnamefont {Gotlieb}}, \bibinfo
  {author} {\bibfnamefont {C.}~\bibnamefont {Hwang}}, \bibinfo {author}
  {\bibfnamefont {D.-H.}\ \bibnamefont {Lee}}, \bibinfo {author} {\bibfnamefont
  {S.~G.}\ \bibnamefont {Louie}}, \bibinfo {author} {\bibfnamefont {J.~D.}\
  \bibnamefont {Denlinger}}, \bibinfo {author} {\bibfnamefont {C.~R.}\
  \bibnamefont {Rotundu}}, \bibinfo {author} {\bibfnamefont {R.~J.}\
  \bibnamefont {Birgeneau}}, \bibinfo {author} {\bibfnamefont {Z.}~\bibnamefont
  {Hussain}}, \ and\ \bibinfo {author} {\bibfnamefont {A.}~\bibnamefont
  {Lanzara}},\ }\href@noop {} {\bibfield  {journal} {\bibinfo  {journal} {Nat.
  Phys.}\ }\textbf {\bibinfo {volume} {9}},\ \bibinfo {pages} {293} (\bibinfo
  {year} {2013})}\BibitemShut {NoStop}%
\bibitem [{\citenamefont {Hsieh}\ \emph {et~al.}(2009)\citenamefont {Hsieh},
  \citenamefont {Xia}, \citenamefont {Qian}, \citenamefont {Wray},
  \citenamefont {Dil}, \citenamefont {Meier}, \citenamefont {Osterwalder},
  \citenamefont {Patthey}, \citenamefont {Checkelsky}, \citenamefont {Ong},
  \citenamefont {Fedorov}, \citenamefont {Lin}, \citenamefont {Bansil},
  \citenamefont {Grauer}, \citenamefont {Hor}, \citenamefont {Cava},\ and\
  \citenamefont {Hasan}}]{HsiehNature_BiSeSpin}%
  \BibitemOpen
  \bibfield  {author} {\bibinfo {author} {\bibfnamefont {D.}~\bibnamefont
  {Hsieh}}, \bibinfo {author} {\bibfnamefont {Y.}~\bibnamefont {Xia}}, \bibinfo
  {author} {\bibfnamefont {D.}~\bibnamefont {Qian}}, \bibinfo {author}
  {\bibfnamefont {L.}~\bibnamefont {Wray}}, \bibinfo {author} {\bibfnamefont
  {J.~H.}\ \bibnamefont {Dil}}, \bibinfo {author} {\bibfnamefont
  {F.}~\bibnamefont {Meier}}, \bibinfo {author} {\bibfnamefont
  {J.}~\bibnamefont {Osterwalder}}, \bibinfo {author} {\bibfnamefont
  {L.}~\bibnamefont {Patthey}}, \bibinfo {author} {\bibfnamefont {J.~G.}\
  \bibnamefont {Checkelsky}}, \bibinfo {author} {\bibfnamefont {N.~P.}\
  \bibnamefont {Ong}}, \bibinfo {author} {\bibfnamefont {A.~V.}\ \bibnamefont
  {Fedorov}}, \bibinfo {author} {\bibfnamefont {H.}~\bibnamefont {Lin}},
  \bibinfo {author} {\bibfnamefont {A.}~\bibnamefont {Bansil}}, \bibinfo
  {author} {\bibfnamefont {D.}~\bibnamefont {Grauer}}, \bibinfo {author}
  {\bibfnamefont {Y.~S.}\ \bibnamefont {Hor}}, \bibinfo {author} {\bibfnamefont
  {R.~J.}\ \bibnamefont {Cava}}, \ and\ \bibinfo {author} {\bibfnamefont
  {M.~Z.}\ \bibnamefont {Hasan}},\ }\href@noop {} {\bibfield  {journal}
  {\bibinfo  {journal} {Nature}\ }\textbf {\bibinfo {volume} {460}},\ \bibinfo
  {pages} {1101} (\bibinfo {year} {2009})}\BibitemShut {NoStop}%
\bibitem [{\citenamefont {Han}\ \emph {et~al.}(2015)\citenamefont {Han},
  \citenamefont {Li}, \citenamefont {Chen}, \citenamefont {Zhu}, \citenamefont
  {Yao}, \citenamefont {Li}, \citenamefont {Wang}, \citenamefont {Gao},
  \citenamefont {Guan}, \citenamefont {Liu}, \citenamefont {Gao}, \citenamefont
  {Qian},\ and\ \citenamefont {Jia}}]{SrBiSeARPESAPL}%
  \BibitemOpen
  \bibfield  {author} {\bibinfo {author} {\bibfnamefont {C.~Q.}\ \bibnamefont
  {Han}}, \bibinfo {author} {\bibfnamefont {H.}~\bibnamefont {Li}}, \bibinfo
  {author} {\bibfnamefont {W.~J.}\ \bibnamefont {Chen}}, \bibinfo {author}
  {\bibfnamefont {F.}~\bibnamefont {Zhu}}, \bibinfo {author} {\bibfnamefont
  {M.-Y.}\ \bibnamefont {Yao}}, \bibinfo {author} {\bibfnamefont {Z.~J.}\
  \bibnamefont {Li}}, \bibinfo {author} {\bibfnamefont {M.}~\bibnamefont
  {Wang}}, \bibinfo {author} {\bibfnamefont {B.~F.}\ \bibnamefont {Gao}},
  \bibinfo {author} {\bibfnamefont {D.~D.}\ \bibnamefont {Guan}}, \bibinfo
  {author} {\bibfnamefont {C.}~\bibnamefont {Liu}}, \bibinfo {author}
  {\bibfnamefont {C.~L.}\ \bibnamefont {Gao}}, \bibinfo {author} {\bibfnamefont
  {D.}~\bibnamefont {Qian}}, \ and\ \bibinfo {author} {\bibfnamefont {J.-F.}\
  \bibnamefont {Jia}},\ }\href@noop {} {\bibfield  {journal} {\bibinfo
  {journal} {Appl. Phys. Lett.}\ }\textbf {\bibinfo {volume} {107}},\ \bibinfo
  {pages} {171602} (\bibinfo {year} {2015})}\BibitemShut {NoStop}%
\bibitem [{\citenamefont {Hor}\ \emph {et~al.}(2010)\citenamefont {Hor},
  \citenamefont {Williams}, \citenamefont {Checkelsky}, \citenamefont
  {Roushan}, \citenamefont {Seo}, \citenamefont {Xu}, \citenamefont
  {Zandbergen}, \citenamefont {Yazdani}, \citenamefont {Ong},\ and\
  \citenamefont {Cava}}]{HorPhysRevLett.104.057001}%
  \BibitemOpen
  \bibfield  {author} {\bibinfo {author} {\bibfnamefont {Y.~S.}\ \bibnamefont
  {Hor}}, \bibinfo {author} {\bibfnamefont {A.~J.}\ \bibnamefont {Williams}},
  \bibinfo {author} {\bibfnamefont {J.~G.}\ \bibnamefont {Checkelsky}},
  \bibinfo {author} {\bibfnamefont {P.}~\bibnamefont {Roushan}}, \bibinfo
  {author} {\bibfnamefont {J.}~\bibnamefont {Seo}}, \bibinfo {author}
  {\bibfnamefont {Q.}~\bibnamefont {Xu}}, \bibinfo {author} {\bibfnamefont
  {H.~W.}\ \bibnamefont {Zandbergen}}, \bibinfo {author} {\bibfnamefont
  {A.}~\bibnamefont {Yazdani}}, \bibinfo {author} {\bibfnamefont {N.~P.}\
  \bibnamefont {Ong}}, \ and\ \bibinfo {author} {\bibfnamefont {R.~J.}\
  \bibnamefont {Cava}},\ }\href@noop {} {\bibfield  {journal} {\bibinfo
  {journal} {Phys. Rev. Lett.}\ }\textbf {\bibinfo {volume} {104}},\ \bibinfo
  {pages} {057001} (\bibinfo {year} {2010})}\BibitemShut {NoStop}%
\bibitem [{\citenamefont {Willa}\ \emph {et~al.}(2018)\citenamefont {Willa},
  \citenamefont {Willa}, \citenamefont {Song}, \citenamefont {Gu},
  \citenamefont {Schneeloch}, \citenamefont {Zhong}, \citenamefont {Koshelev},
  \citenamefont {Kwok},\ and\ \citenamefont
  {Welp}}]{KristinSrBiSeARSHPhysRevB.98.184509}%
  \BibitemOpen
  \bibfield  {author} {\bibinfo {author} {\bibfnamefont {K.}~\bibnamefont
  {Willa}}, \bibinfo {author} {\bibfnamefont {R.}~\bibnamefont {Willa}},
  \bibinfo {author} {\bibfnamefont {K.~W.}\ \bibnamefont {Song}}, \bibinfo
  {author} {\bibfnamefont {G.~D.}\ \bibnamefont {Gu}}, \bibinfo {author}
  {\bibfnamefont {J.~A.}\ \bibnamefont {Schneeloch}}, \bibinfo {author}
  {\bibfnamefont {R.}~\bibnamefont {Zhong}}, \bibinfo {author} {\bibfnamefont
  {A.~E.}\ \bibnamefont {Koshelev}}, \bibinfo {author} {\bibfnamefont {W.-K.}\
  \bibnamefont {Kwok}}, \ and\ \bibinfo {author} {\bibfnamefont
  {U.}~\bibnamefont {Welp}},\ }\href {\doibase 10.1103/PhysRevB.98.184509}
  {\bibfield  {journal} {\bibinfo  {journal} {Phys. Rev. B}\ }\textbf {\bibinfo
  {volume} {98}},\ \bibinfo {pages} {184509} (\bibinfo {year}
  {2018})}\BibitemShut {NoStop}%
\bibitem [{\citenamefont {Sakakibara}\ \emph {et~al.}(2016)\citenamefont
  {Sakakibara}, \citenamefont {Kittaka},\ and\ \citenamefont
  {Machida}}]{SakakibaraReview}%
  \BibitemOpen
  \bibfield  {author} {\bibinfo {author} {\bibfnamefont {T.}~\bibnamefont
  {Sakakibara}}, \bibinfo {author} {\bibfnamefont {S.}~\bibnamefont {Kittaka}},
  \ and\ \bibinfo {author} {\bibfnamefont {K.}~\bibnamefont {Machida}},\
  }\href@noop {} {\bibfield  {journal} {\bibinfo  {journal} {Rep. Prog. Phys.}\
  }\textbf {\bibinfo {volume} {79}},\ \bibinfo {pages} {094002} (\bibinfo
  {year} {2016})}\BibitemShut {NoStop}%
\bibitem [{\citenamefont {Sun}\ \emph {et~al.}(2017)\citenamefont {Sun},
  \citenamefont {Kittaka}, \citenamefont {Nakamura}, \citenamefont
  {Sakakibara}, \citenamefont {Irie}, \citenamefont {Nomoto}, \citenamefont
  {Machida}, \citenamefont {Chen},\ and\ \citenamefont
  {Tamegai}}]{YueSunPRBFeSeARSH}%
  \BibitemOpen
  \bibfield  {author} {\bibinfo {author} {\bibfnamefont {Y.}~\bibnamefont
  {Sun}}, \bibinfo {author} {\bibfnamefont {S.}~\bibnamefont {Kittaka}},
  \bibinfo {author} {\bibfnamefont {S.}~\bibnamefont {Nakamura}}, \bibinfo
  {author} {\bibfnamefont {T.}~\bibnamefont {Sakakibara}}, \bibinfo {author}
  {\bibfnamefont {K.}~\bibnamefont {Irie}}, \bibinfo {author} {\bibfnamefont
  {T.}~\bibnamefont {Nomoto}}, \bibinfo {author} {\bibfnamefont
  {K.}~\bibnamefont {Machida}}, \bibinfo {author} {\bibfnamefont
  {J.}~\bibnamefont {Chen}}, \ and\ \bibinfo {author} {\bibfnamefont
  {T.}~\bibnamefont {Tamegai}},\ }\href {\doibase 10.1103/PhysRevB.96.220505}
  {\bibfield  {journal} {\bibinfo  {journal} {Phys. Rev. B}\ }\textbf {\bibinfo
  {volume} {96}},\ \bibinfo {pages} {220505} (\bibinfo {year}
  {2017})}\BibitemShut {NoStop}%
\bibitem [{\citenamefont {Hecker}\ and\ \citenamefont
  {Schmalian}(2018)}]{HeckerNPJQM}%
  \BibitemOpen
  \bibfield  {author} {\bibinfo {author} {\bibfnamefont {M.}~\bibnamefont
  {Hecker}}\ and\ \bibinfo {author} {\bibfnamefont {J.}~\bibnamefont
  {Schmalian}},\ }\href@noop {} {\bibfield  {journal} {\bibinfo  {journal} {npj
  Quantum Materials}\ }\textbf {\bibinfo {volume} {3}},\ \bibinfo {pages} {26}
  (\bibinfo {year} {2018})}\BibitemShut {NoStop}%
\end{thebibliography}%

\clearpage
\onecolumngrid
\begin{center}
	\textbf{\huge Supplemental information}
\end{center}
\vspace{1cm}
\twocolumngrid
\setcounter{equation}{0}
\setcounter{figure}{0}
\setcounter{table}{0}
\setcounter{page}{1}
\makeatletter
\renewcommand{\theequation}{S\arabic{equation}}
\renewcommand{\thefigure}{S\arabic{figure}}

\section*{S1 Experiment system and the calibration}

\begin{figure}\center
	\includegraphics[width=8.5cm]{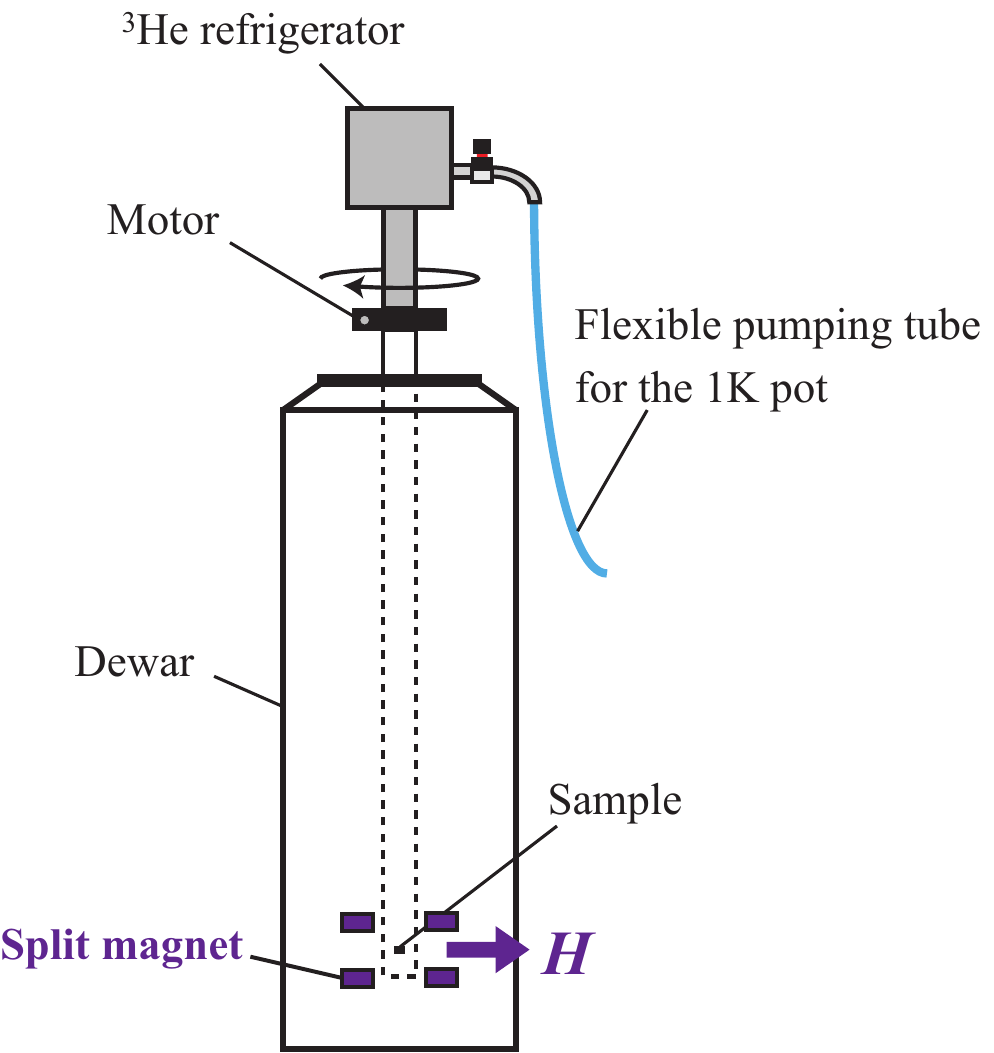}\\
	\caption{Schematic drawing of the angle-resolved specific heat measurement system}\label{}
\end{figure}

\begin{figure}\center
	\includegraphics[width=8.5cm]{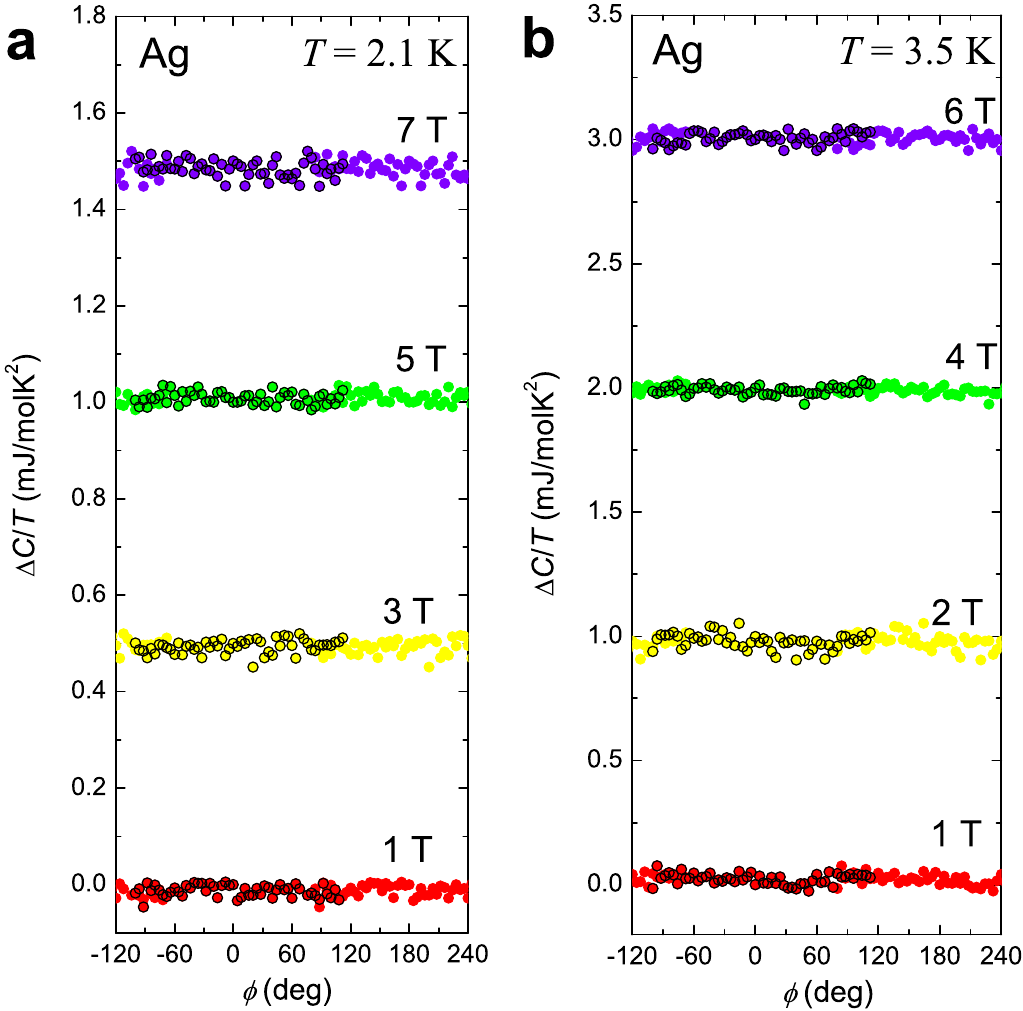}\\
	\caption{Azimuthal angle dependence of the specific heat $\Delta C(\phi)/T$ of Ag measured under various fields at (a) 2.1 K, and (b) 3.5 K. Here, $\Delta C(\phi)/T$ is defined as $C(\phi)/T$-$C(0^\circ)/T$, and each curve is shifted vertically by 0.5 mJ/molK$^2$ for (a), and 1 mJ/molK$^2$ for (b), respectively. Symbols with black border are the measured data, and those without black border is mirrored points to show the symmetry more clearly.}\label{}
\end{figure}

Details about the angle-resolved specific heat (ARSH) measurement system is shown schematically in Fig.S1. A $^3$He refrigerator (Heliox, Oxford) was used to cool the samples to temperatures below 300 mK. Magnetic fields were generated by a split-pair superconducting magnet, capable of producing 8 T horizontal fields. Having no thick pumping tube outside, the refrigerator insert could be easily rotated using a stepper motor mounted at the top of a magnet Dewar. The overall angular resolution of the field direction is better than 0.01 deg. The hexagonal plan of the sample can be set either parallel or perpendicular (with a rectangular accessory) to the horizontal magnetic field to measure the azimuthal angle ($\phi$) or the polar angle ($\theta$) dependences of the specific heat. During the measurements, the angles obtained from the number of pulse to the motor were checked equal to those obtained from the goniometer on the motor, which confirms that the correct angles were obtained. 

To confirm that the our measurement system contains no field-angle dependent background, we performed the ARSH measurements on a piece of Ag with almost the same weight as the Sr$_x$Bi$_2$Se$_3$ sample. In such experiment, the whole system is the same as that used for the sample, and even the gravitational force to the addenda is nearly the same. Some typical results at 2.1 K and 3.5 K are shown in Fig. S2. Clearly, no oscillation of ARSH is observed, which proves that there is no field-angle dependent background of the system. Thus, the two-fold symmetric ARSH results observed in Sr$_x$Bi$_2$Se$_3$ is the intrinsic property of the sample.

\section*{S2 Azimuthal angle dependence of the specific heat $\Delta C(\phi)/T$ at 3.5 K and 5 K}

\begin{figure}\center
	\includegraphics[width=8.5cm]{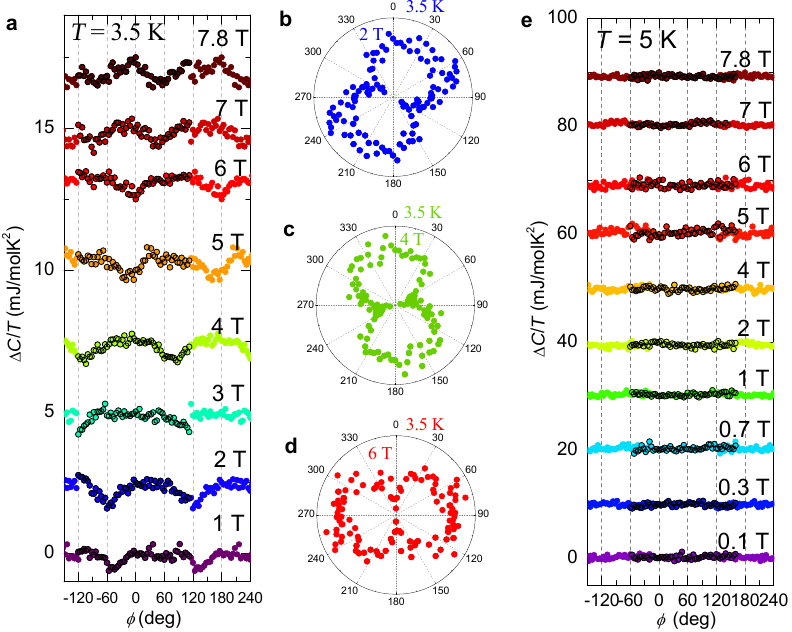}\\
	\caption{(a) Azimuthal angle dependence of the specific heat $\Delta C(\phi)/T$ measured under various fields at 3.5 K. $\Delta C(\phi)/T$ is defined as $C(\phi)/T$-$C(0^\circ)/T$, and each subsequent curve is shifted vertically by 2.5 mJ/molK$^2$. Symbols with black outlines are measured data, and those without are mirrored points to show the symmetry. (b)-(d) Polar plot of the $\Delta C(\phi)/T$ at 3.5 K under fields of 2 T, 4 T and 6 T, respectively. (e) Azimuthal angle dependence of the specific heat $\Delta C(\phi)/T$ measured under various fields at 5 K. $\Delta C(\phi)/T$ is defined as $C(\phi)/T$-$C(0^\circ)/T$, and each subsequent curve is shifted vertically by 10 mJ/molK$^2$. }\label{}
\end{figure}

Figure S3(a) shows the azimuthal angle dependence of the specific heat $\Delta C(\phi)/T$ measured at 3.5 K under different fields. Since 3.5 K is already above $T_{\rm{c}}$ ($\sim$ 3.1 K), the measurements probe the normal state properties of the Sr$_x$Bi$_2$Se$_3$. Obviously, $\Delta C(\phi)/T$ manifests two-fold symmetry, which can be seen more clear in the polar plot shown in Figs. S3(b)-(d) for selected data at 2, 4, and 6 T. Figs. S3(e) shows the azimuthal angle dependence of the specific heat $\Delta C(\phi)/T$ measured at 5 K under different fields. Obviously, the two-fold symmetric oscillation disappears at 5 K at least in the resolution limit of our measurement system. We also want to remind that at higher temperature, the phonon contribution is much larger ($C_{phonon}$ $\sim$ $T^3$), some tiny oscillation may be covered due to the larger background.   

\section*{S3 Polar angle dependence of the specific heat $\Delta C(\theta)/T$}

\begin{figure}\center
	\includegraphics[width=8.5cm]{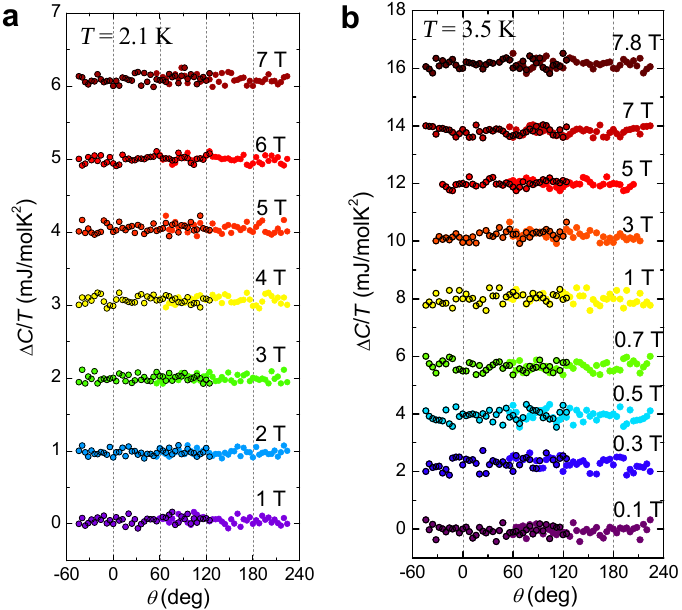}\\
	\caption{Polar angle dependence of the specific heat $\Delta C(\theta)/T$ measured under different fields at (a) 2.1 K, and (b) 3.5 K. $\Delta C(\theta)/T$ is defined as $C(\theta)/T$-$C(0^\circ)/T$, and each subsequent curve is shifted vertically by 1 mJ/molK$^2$ in (a), and 2 mJ/molK$^2$ in (b), respectively. Black-outlined symbols are the measured data; the others are mirrored points to show the symmetry more clearly.}\label{}
\end{figure}

Figure S4 shows the polar angle dependence of the specific heat $\Delta C(\theta)/T$ measured at 2.1 K and 3.5 K under different fields. During the polar angle dependent measurements, the azimuthal angle was kept as $\phi$ = 0$^\circ$. Obviously, the specific heat result at fixed magnetic field is independent of the out-of-plane angle $\theta$, which proves that the two-fold symmetric $\Delta C(\phi)/T$ results could not be attributed to the $\Delta C(\theta)/T$ signal from the misalignment of the sample setting. 

\section*{S4 Schottky anomaly at low temperature}

\begin{figure}\center
	\includegraphics[width=8.5cm]{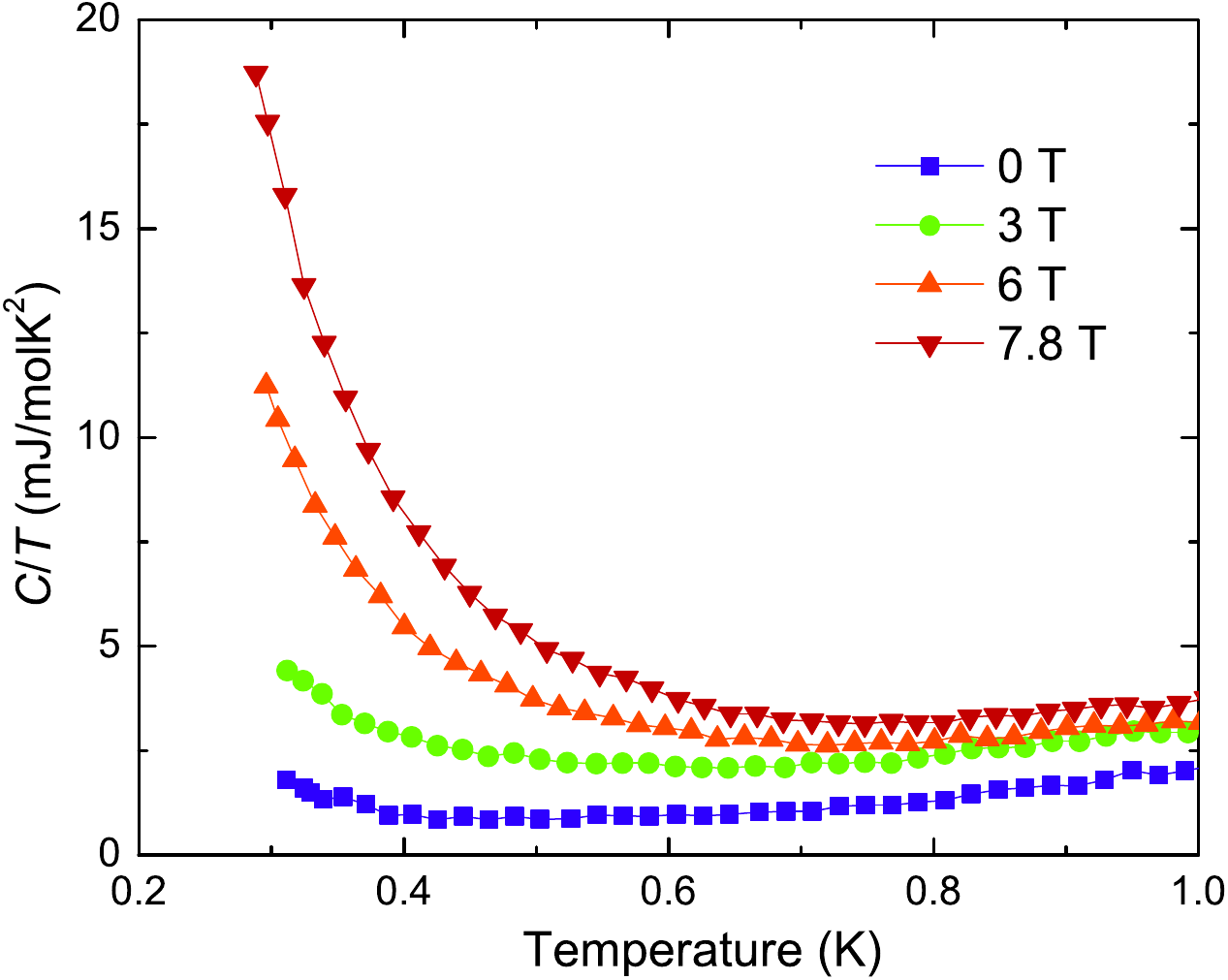}\\
	\caption{Temperature dependence of specific heat $C$/$T$ at temperatures below 1 K, measured under magnetic fields of 0, 3 T, 6 T, and 7.8 T. }\label{}
\end{figure}

The temperature dependence of specific heat at low temperatures under magnetic fields of 0, 3 T, 6 T, and 7.8 T is shown in Fig. S5. Under zero field, a trace of upturn can be observed at temperatures below 0.4 K. With increasing the field, the upturn in specific heat becomes stronger, however, it is only observed below 0.7 K even under 7.8 T. Such upturn is attributed to the Schottky behavior of nuclear contributions, mainly from the Bi with nuclear spin $I$ = 9/2. Similar behavior is also observed in Cu$_x$Bi$_2$Se$_3$ \cite{YonezawaARSHCuBiSe}. Since the Schottky anomaly is only observed at temperatures below 0.8 K, which will not affect the conclusion of our paper.

\section*{S5 Superconducting jump in specific heat $\Delta C$/$T_{\rm{c}}$}

Figure S6(a) shows the specific heat divided by temperature $C$/$T$ as a function of $T^2$ measured under zero-filed. A jump attributed to the SC transition can be clearly witnessed although the magnitude is very small. Similar small SC jump in the specific heat of M$_x$Bi$_2$Se$_3$ is also reported previously \cite{YonezawaARSHCuBiSe,KristinSrBiSeARSHPhysRevB.98.184509}. After subtracting the normal state specific heat obtained simply by linear fitting of the data above $T_{\rm{c}}$ ($C/T$ = -3.94 + 4.18$T^2$), the specific heat jump $\Delta C$/$T$ is obtained and shown in Fig. S6(b). The SC jump $\Delta C$/$T_{\rm{c}}$ is estimated as $\sim$ 1.1 mJ/mol K$^2$, which is also similar to previous reports \cite{YonezawaARSHCuBiSe,KristinSrBiSeARSHPhysRevB.98.184509}.

\begin{figure}\center
	\includegraphics[width=8.5cm]{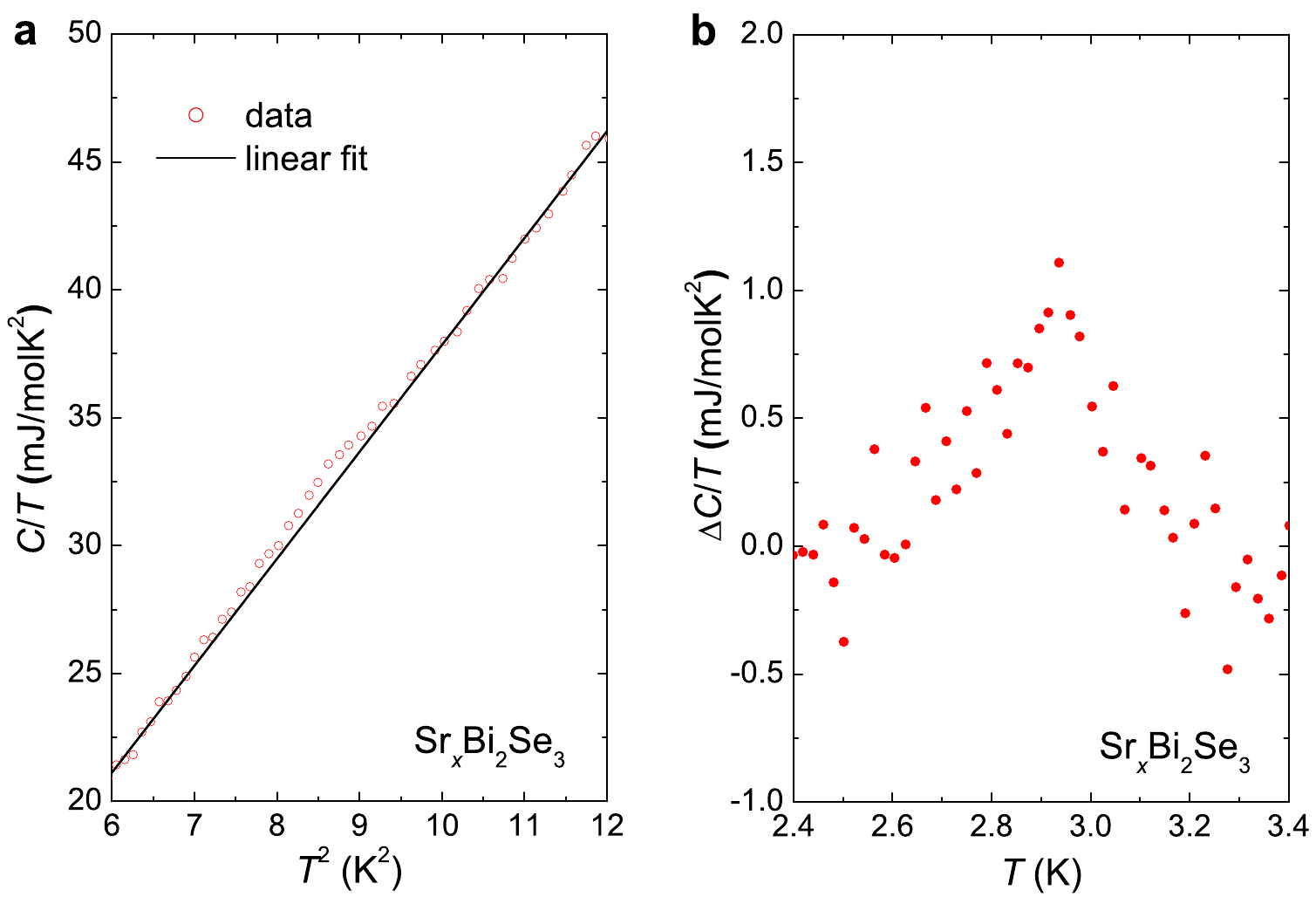}\\
	\caption{(a) Raw data of the specific heat divided by temperature $C$/$T$, including the contributions from phonon and addenda as a function of $T^2$ measured under zero-filed. The solid line represent the linear fit to the normal state specific heat. (b) Temperature dependence of the specific heat $\Delta C$/$T$ obtained by subtracting the fitting result of the normal state specific heat.}\label{}
\end{figure}

\section*{S6 Discussion of the multi-domain effect and the change of the symmetry axis by magnetic field}
The multi-domain effect is a common feature of materials with spontaneously rotational symmetry breaking, which has already been observed in M$_x$Bi$_2$Se$_3$ \cite{YonezawaARSHCuBiSe,KuntsevichNJP}. To directly show the effect of multi-domain effect to the angle-dependent specific heat, we consider a simple situation of three domains (labeled domain A, domain B, and domain C). The angles between every two domains are 60$^\circ$. The proportion of the three domains are 20\%, 30\%, and 50\% for domain A, B, and C, respectively (see Fig. S7(a)). Similar assumption was also applied to the discussion of multi-domain effect in Cu$_x$Bi$_2$Se$_3$ (Supplementary S7 of Ref. \cite{YonezawaARSHCuBiSe}). The dashed lines show the 2-fold symmetric waveforms for the three domains. Their superposition is plotted by the solid line, which shows 2-fold symmetry (see Fig. S7(e)). Except for the special case that the magnitude of the three oscillations is just canceled, the total wave keeps 2-fold symmetry. It confirms that the multi-domain effect will not affect the conclusion of the 2-fold symmetry. Of course, too many randomly distributed fine domains will make the symmetry almost canceled, and there will be no symmetric signal, which is obvious different from our observations. 

Then, we consider the field-induced symmetric axis rotation in a single domain as observed by the STM measurements \cite{TaoRanPhysRevX.8.041024}. Fig. S7(b) shows the situation that symmetric axes of all the three domains spontaneously rotate 30$^\circ$ in clockwise. The total waveform also shifts 30$^\circ$, while keeps the same amplitude (see Fig. S7(f); Here, we do not consider the increase of magnitude due to the Zeeman effect.). Under larger magnetic field, the domains are expected to rotate to larger angles. Thus, the symmetric axis will continually rotates with increasing field. If we include the increase of Zeeman effect with increasing field, the magnitude of the 2-fold oscillation will also be enhanced. It is similar to the observation of the angle-dependent specific heat at 0.35 K (Fig. 2(a)). On the other hand, if only the symmetric axis of one domain rotates (see the examples of domain C rotates 30$^\circ$ in Fig. S7(c), and domain B rotates 30$^\circ$ in (d)), not only the symmetric axis shifts, but also the amplitude of the total waveform is changed (see Fig. S7(g) and (h)). It can explain the irregular changes in the symmetric axis and the magnitude of the oscillation at higher temperatures around $T_{\rm{c}}$. Considering the theory of Ref.\cite{HeckerNPJQM}, the nematic phase is originated from the superconducting fluctuation, and its critical temperature is comparable to $T_{\rm{c}}$. The nematic state around $T_{\rm{c}}$ is very sensitive to the magnetic field and temperature, and may be fluctuated between domains. In this situation, the symmetry axis change only happens in some domains as shown in Fig. S7(c)-(d)). On contrary, at low temperatures far below $T_{\rm{c}}$, the nematic order is strong enough that all the domains show the same behavior under the same fields or temperatures.

\begin{figure}\center
	\includegraphics[width=8.5cm]{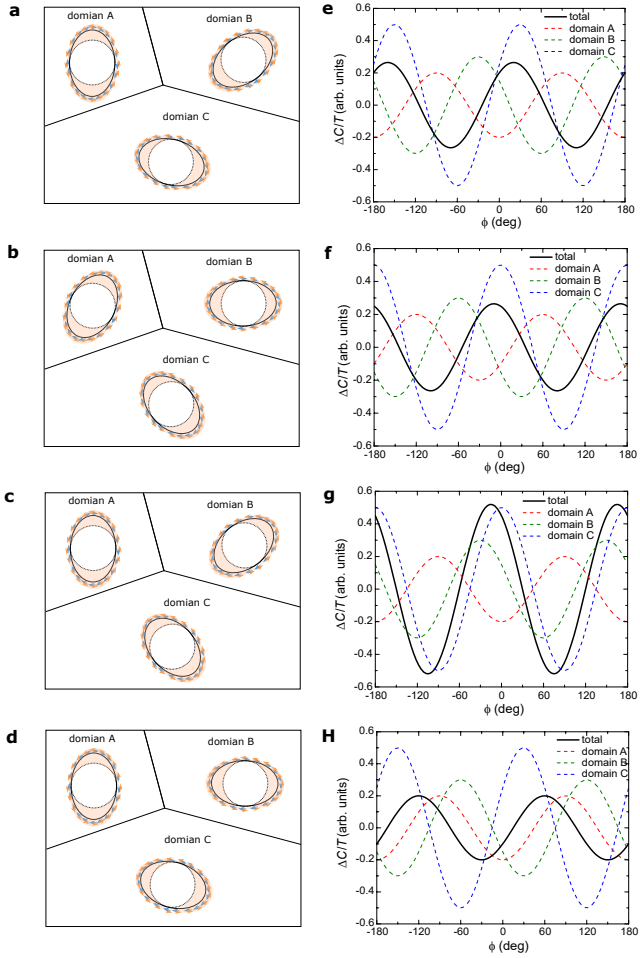}\\
	\caption{Schematic drawing of (a) three domains in the crystal. (b) all the three domains rotates 30$^\circ$ spontaneously. (c) only the domain C rotates 30$^\circ$. (d) only the domain D rotates 30$^\circ$. (e)-(h) the corresponding angle-dependent specific heat ($\Delta C/T$ in the arbitrary unity) of (a)–(d), respectively. The dashed lines represent the result of each domain, while the solid line represents the total result of the superposition of the three domains.}\label{}
\end{figure}


\end{document}